\documentclass[journal,draft,onecolumn]{IEEEtran}
\usepackage{cite}
\usepackage{graphicx}
\usepackage{amsmath}
\usepackage{array}
\usepackage{enumerate}
\usepackage{amsbsy}
\usepackage{amssymb}
\usepackage{amsthm}
\usepackage{amscd}
\usepackage{subfigure}
\usepackage{float}
\usepackage{balance}

\usepackage{amsmath}
\usepackage{algpseudocode}
\usepackage{amsthm}
\usepackage{dsfont}
\usepackage{algorithm}
\usepackage{thmtools}
\usepackage{amssymb}
\usepackage[dvipsnames]{xcolor}
\usepackage{todonotes}

\def\bR{\boldsymbol{R}}

\def\mbf{\mathbf{f}}

\def\mbx{\mathbf{x}}
\def\mby{\mathbf{y}}
\def\mbz{\mathbf{z}}

\def\mbB{\mathbf{B}}
\def\mbC{\mathbf{C}}
\def\mbD{\mathbf{D}}

\def\mbF{\mathbf{F}}

\def\mbI{\mathbf{I}}

\def\mbR{\mathbf{R}}

\def\mbY{\mathbf{Y}}

\def\bzero{\boldsymbol{0}}
\def\bone{\boldsymbol{1}}

\newcommand{\complexC}[1][]{\mathds{C}^{#1}}

\newcommand{\realR}[1][]{\mathds{R}^{#1}}

\def\Diag#1{\mathrm{Diag}\left(#1\right)}

\newcommand{\zinv}[1][]{z^{\ifx&#1&-1\else-#1\fi}}

\def\Oh#1{\mathcal{O}\left(#1\right)}

\def\st{~\mathrm{s.t.}~}

\def\ew{~\mathrm{elsewhere}}
\def\eg{\textit{e.g.}}
\def\ie{\textit{i.e.}}

\def\blue#1{\textcolor{blue}{#1}}

\theoremstyle{definition}

\declaretheorem[style=definition,name=Remark,qed=$\blacksquare$]{remark}

\usepackage{xpatch}
\makeatletter
\xpatchcmd{\@thm}{\thm@headpunct{.}}{\thm@headpunct{}}{}{}
\makeatother

\algnewcommand\algorithmicinput{\textbf{Input:}}
\algnewcommand\Input{\item[\algorithmicinput]}
\algnewcommand\algorithmicoutput{\textbf{Output:}}
\algnewcommand\Output{\item[\algorithmicoutput]}
\algnewcommand\algorithmicinit{\textbf{Initialize:}}
\algnewcommand\Init{\item[\algorithmicinit]}

\def\To{\textbf{to}~}

\title{Waveform Design for Mutual Interference Mitigation in Automotive Radar}

\author{Arindam~Bose\IEEEauthorrefmark{1},~\IEEEmembership{Student Member,~IEEE}, Bo Tang\IEEEauthorrefmark{1}, Wenjie Huang,~\IEEEmembership{Student Member,~IEEE}\\ Mojtaba~Soltanalian\IEEEauthorrefmark{2},~\IEEEmembership{Senior Member,~IEEE}, Jian Li\IEEEauthorrefmark{2},~\IEEEmembership{Fellow,~IEEE}
\thanks{\IEEEauthorrefmark{1}The authors contributed equally to this work. \IEEEauthorrefmark{2}Corresponding authors.}
\thanks{This work was supported in part by U.S. National Science Foundation Grants ECCS-1708509 and ECCS-1809225, and an Illinois Discovery Partners Institute (DPI) Seed Award. Parts of this work have been presented at the 2018 IEEE International Conference on Acoustics, Speech and Signal Processing (ICASSP), Calgary, AB, Canada in April, 2018 \cite{8461806}.}
\thanks{A. Bose and M. Soltanalian are with the Department of Electrical and Computer Engineering, University of Illinois at Chicago, Chicago, IL 60607 USA (e-mail: abose4, msol@uic.edu).}
\thanks{B. Tang is with the Department of Electronic Engineering and Information Science, University of Science and Technology of China, Hefei 230027, China, and also with the Electronic Engineering Institute, Hefei 230000, China (e-mail: tangbo06@gmail.com).}
\thanks{W. Huang is with the Department of Electronic Engineering and Information Science, University of Science and Technology of China, Hefei, China.}
\thanks{J. Li is with the Department of Electrical and Computer Engineering, University of Florida, Gainesville, FL 32611 USA (e-mail: li@dsp.ufl.edu).}}

\begin{document}
\maketitle
\begin{abstract}
The mutual interference between similar radar systems can result in reduced radar sensitivity and increased false alarm rates. 
To address the synchronous and asynchronous interference mitigation problems in similar radar systems, we first propose herein two slow-time coding schemes to modulate the pulses within a coherent processing interval (CPI) for a single-input-single-output (SISO) scenario. Specifically, the first coding scheme relies on Doppler shifting and the second one is devised based on an optimization approach. 
We further extend our discussion to the more general case of multiple-input-multiple-output (MIMO) radars and propose an efficient algorithm to design waveforms to mitigate mutual interference in such systems.
The proposed coding schemes are computationally efficient in practice and the incorporation of the coding schemes requires only a slight modification of the existing systems.
Our numerical examples indicate that the proposed coding schemes can reduce the interference power level in a desired area of the cross-ambiguity function significantly.
\end{abstract}

\begin{IEEEkeywords}
	Automotive radar systems, mutual interference mitigation, slow-time coding, code optimization, MIMO.
\end{IEEEkeywords}

\IEEEpeerreviewmaketitle

\section{Introduction}
\label{sec:intro}

\IEEEPARstart{T}{he} radar technology exhibits an unmatched performance in a variety of vehicular applications, owing to its excellent target resolving capabilities in bad weather and low visibility conditions in comparison with visible and infrared imaging techniques~\cite{7870764, Schneider05automotiveradar, 5547432, 7485214, gini2012waveform}.
In recent years, due to the benefits of exploiting the millimeter-wave (mm-wave) band such as increased bandwidth, high spatial resolution, smaller size and weight of mm-wave equipment compared to ultrasonic radar and lidar, the mm-wave radars (30-300 GHz) have gained significant popularity not only in automotive radar systems but also in drone radar applications and internet of things (IoT) \cite{10.1145/2897824.2925953, 10.1145/2984511.2984565, saponara2019radar}.
For instance, mm-wave radar equipment is much simpler than lidar's complex mirrors and lasers, but far faster and more accurate than ultrasonic radars, and may use highly directional, high-frequency EM waves to map the surrounding environment \cite{greco2019advances1, greco2019advances2}.
Given the tendency to mass-produce radars for civilian applications (\eg, automotive radars), such systems, however, tend to be quite similar, or even almost identical. The increasing number of similar or identical radar systems increases the probability of mutual interference, which may result in severely reduced radar sensitivity and poor performance quality \cite{7870737, 8828025, khobahi2019deep}.
Thus it is vitally important to enhance radar signal processing performance in severe mutual interference scenarios
\cite{8642926, 9048834, 9054680, AlaeeKerahroodi2019CDMMIMOIN}.

In the literature, the effects of mutual interference and their corresponding methods of mitigation have been discussed widely, \eg, see \cite{8555539, Skaria2019, rs6010740, 2010AdRS855G, 5760761, 7944482, 8079965, 9127843, 4106078, 9266283, 8400806, 9053892, 6334043} and the references within.
A judicious signal separation method for synchronous and asynchronous interference mitigation is proposed in \cite{8555539}.
The authors in \cite{Skaria2019} study the impact of radar waveform design and the associated receiver processing on the statistics of radar-to-radar interference and further propose an approach based on pseudo-random cyclic orthogonal sequences (PRCOS), which enable sensors to rapidly learn the interference environment and avoid using frequency overlapping waveforms.
An iterative filtering algorithm followed by matched filtering for each radar has been suggested in \cite {rs6010740} to suppress the radar-to-radar interferences.
Furthermore, in \cite{4106078}, the authors analyzed the mutual interference between frequency-modulated continuous-wave (FMCW) radar systems and proposed several techniques to mitigate the interference problem, including pre-possessing and finite impulse response (FIR) filtering.
Contrary to addressing the mutual interference in the receiver side, the authors in \cite{2010AdRS855G, 5760761, 7944482} investigated the problems between automotive radar systems with different types of transmissions.

In this paper, we address the mutual interference mitigation problem for similar or identical radar systems by using smart transmit waveforms.
In particular, we consider a scenario where the manufacturer of the radar systems present in the scene is essentially the same; as a result, the radar parameters such as carrier frequency, slope, and the number of the chirps are assumed to be similar. 
It is true that in the real world, such a scenario is currently relatively rare. However, as more cars are equipped with radars, this may occur more frequently; something that can be dealt with through future regulations and shared protocols.
We further assume that in such radar systems, it is possible to modify the custom waveforms used for transmission \textit{on-the-fly}.
Note that such a scenario can be avoided simply by ``orthogonalization'' of radar parameters. Designing waveforms, however, can make such a process more optimal, particularly when more than two vehicles are involved. In this sense, this work lays the ground for future sensible solutions in multi-vehicle scenarios.

To this end, we first formulate the problem of waveform designing for a simple single-input-single-output (SISO) case.
Particularly, we propose two slow-time coding schemes to reduce the interference power level. 
Next, we extend our approach to the more general multiple-input-multiple-output (MIMO) scenario.
Recently, there has been considerable interest in radar systems employing multiple antennas at both the transmitter and receiver; thus, performing space-time processing on both~\cite{4383615}.
Compared with the traditional phased-array radars, a MIMO radar can transmit multiple independent probing waveforms simultaneously, enabling further waveform diversity \cite{9050863} and enhanced parameter identifiability \cite{4358016}.
Given the pervasive use of MIMO systems, it is of utter importance to design sets of waveforms that will reduce mutual interference between different MIMO automotive radars. 
The main contributions of this work can be summarized as follows:\vspace{3pt}
\begin{itemize}
    \item We begin our study with the SISO case and propose two coding schemes to reduce the interference power level. The first coding scheme aims to shift the Doppler frequency of the interference and separate it from the target in the Doppler region. The second coding scheme aims to minimize the discrete periodic cross-ambiguity function (PCAF) in the desired area. \vspace{5pt}
    
    \item We further formulate the problem of minimizing the discrete PCAF for a general MIMO scenario and propose an efficient cyclic algorithm to design the transmit waveforms. 
\end{itemize}
Note that a distributed and mutually cooperative design protocol can be adopted for the online design of such waveforms in a collaborative manner. The design can also be done offline where the radar codes can be designed and stored in a centralized radar codebook for later usage.
The discussion of such design coordination protocols is, however, beyond the scope of this work.

The rest of this paper is organized as follows. 
In Section \ref{sec:pre}, we discuss the mutual interference for two identical frequency modulated continuous wave (FMCW) radar systems.
We propose two coding schemes for the SISO scenario in Section~\ref{sec:siso}.
Section~\ref{sec:mimo} is devoted to waveform design for the more general MIMO scenario. Numerical simulation results are presented in Section \ref{sec:num}.
Finally, Section \ref{sec:con} concludes our work.
\vspace{5pt}

\textit{Notation:} We use bold-lowercase and bold-uppercase letters to represent vectors and matrices, respectively. 
$x_i$ denotes the $i$-th element of the vector $\mathbf{x}$. The superscripts $(\cdot)^*$, $(\cdot)^T$, and $(\cdot)^H$ represent the conjugate, the transpose, and the Hermitian operators, respectively.
The set of complex matrices are denoted by $\complexC$. 
$\Re\{\cdot\}$ signifies the real part of a complex variable.
The $\ell_p$-norm is represented by $\|\cdot\|_p$. The  Frobenius norm of the matrix $\mathbf{X}$ is denoted by $\|\mathbf{X}\|_F $. 
$\mathbf{I}_N$ is the identity matrix of size $N\times N$, while $\mathbf{0}$ denotes the all-zero vector/matrix of given size.
$\Diag{\mathbf{x}}$ denotes a diagonal matrix formed whose diagonal elements are determined by the vector $\mathbf{x}$.
Finally, $\odot$ represents the Hadamard product of matrices.

\nocite{8926388, 8314765}

\section{Preliminaries} \label{sec:pre}
\begin{figure}[!t]
  \centering
  {\subfigure[]{{\includegraphics[width = 0.6\textwidth,draft=false]{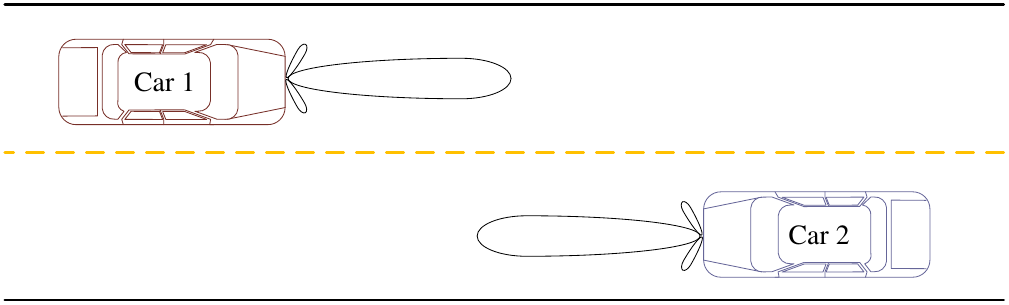}} \label{fig:1a}} }
  {\subfigure[]{{\includegraphics[width = 0.6\textwidth,draft=false]{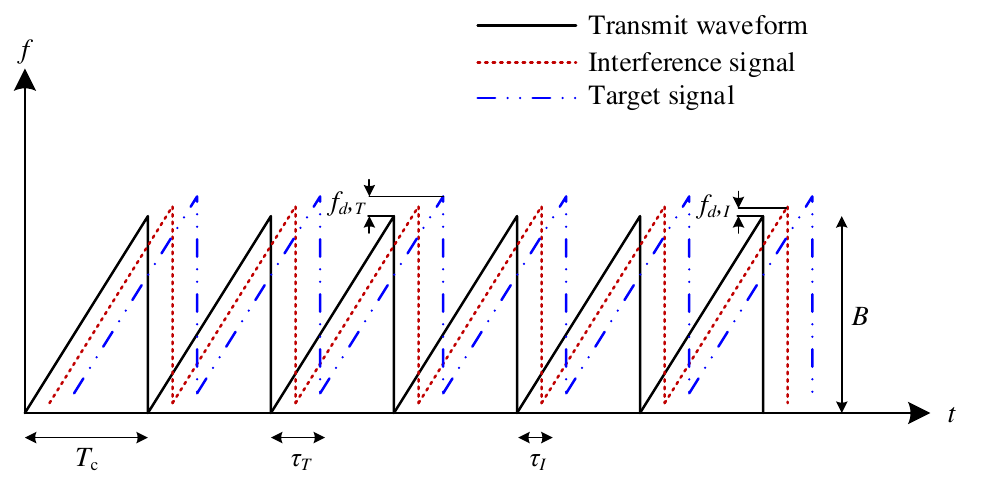}} \label{fig:1b}} }
  \caption{Mutual interference between automotive radars: (a) potential source of mutual interference; (b) time-frequency illustration of the transmit waveform, the target signal, and the interference.}\label{fig:1}
\end{figure}

In this section, we start by formulating the problem for a SISO scenario.
Consider two identical FMCW radar systems shown in Fig. \ref{fig:1a}, that are operating within the same frequency band and same $B/T_c$ ratio where $B$ and $T_c$ are the FMCW signal bandwidth and chirp time, respectively, as shown in Fig. \ref{fig:1b}. The transmitted signal can be expressed as
\begin{align}
    s(t)=\sum_{n=0}^{C-1}{u(t-nT_c)}
\end{align}
where $u(t)=\frac{1}{\sqrt{T_c}}\exp\left(j(2\pi f_c t + \pi Kt^2)\right)\textrm{rect}\left(\frac{t-T_c}{T_c}\right)$, $f_c$ is the carrier frequency, $C$ is the number of chirps, $K=B/T_c$, and
\begin{align*}
    \textrm{rect}(t) = \left\lbrace \begin{array}{ll}
         1, & 0 \leq t \leq 1, \\
         0, & \ew.
    \end{array}\right.
\end{align*}

When the two radar systems are operating simultaneously, the received signal by one radar includes not only the target reflections but also the interference signal due to the transmission from the other radar system. As a result, we can express the received signal by one radar (\eg, the radar mounted on Car 1 in Fig. \ref{fig:1a}) as,
\begin{align}
r(t) = y_T(t)+y_I(t)+w(t)    
\end{align}
where
$y_T(t)=\alpha_Ts(t-\tau_T)\exp(j\pi f_{d,T}t)$ is the target return, and 
$y_I(t)=\alpha_Is(t-\tau_I)\exp(j\pi f_{d,I}t)$ is the interference signal with 
$\alpha_T, \alpha_I$ being the corresponding amplitudes,
$\tau_T$ is the two-way target propagation delay,
$\tau_I$ is the one-way delay associated with the interference,
$f_{d,T}, f_{d,I}$ are the corresponding Doppler frequencies, and
$w(t)$ is the internal disturbance, including, \eg, the receiver noise.

Typically, FMCW radar systems collect the received signal from $N$ consecutive pulses within a coherent processing interval (CPI) for target detection and parameter estimation. The received signal is then conjugately mixed with the transmitted signal to produce a low-frequency intermediate (de-chirped) signal.
As a result, the de-chirped version of $r(t)$ for the $n^{th}$ (slow-time) pulse is given by
\begin{align}
    r_{dc}^n(t) &= \alpha_T\exp(j2\pi(f_{B,T}t+nf_{d,T}T_c)) \nonumber \\ &~+\alpha_I\exp(j2\pi(f_{B,I}t+nf_{d,I}T_c)) + w^n(t)
\end{align}
where 
$f_{B,T} = K{\tau_T}+f_{d,T}$,
$f_{B,I} = K{\tau_I}+f_{d,I}$
are the beat frequencies corresponding to the target and the interference signal, respectively, and to lighten the notations, we absorb the constant phase terms into $\alpha_T$ and $\alpha_I$, and use $w^n(t)$ to denote the de-chirped noise.

The de-chirped signal is then passed through analog-to-digital converters (ADCs) and the $m^{th}$ (fast-time) digital sample can be expressed as,
\begin{align}\label{eq:rmn}
  r(m,n) &= \alpha_T\exp(j2\pi(\hat{f}_{B,T}m+\hat{f}_{d,T}n)) \nonumber \\ &+ \alpha_I\exp(j2\pi(\hat{f}_{B,I}m+\hat{f}_{d,I}n)) + w(m,n)
\end{align}
where
$\hat{f}_{B,T}=f_{B,T}T_s,~\hat{f}_{B,I}=f_{B,I}T_s$ are the corresponding normalized beat frequencies,
$\hat{f}_{d,T}=f_{d,T}T_c, ~\hat{f}_{d,I}=f_{d,I}T_c$ are the corresponding normalized Doppler frequencies, and
$T_s = 1/f_s$, with $f_s$ being the sampling frequency.

Applying 2-D FFT to \eqref{eq:rmn} for $m=1,\cdots,M$ and $n=1,\cdots,N$, we can obtain the range-Doppler image as,
\begin{align}\label{Eq:RDmap}
    \text{RD}(k,p) &= \alpha_TD_M(\hat{f}_{B,T}-k/M) D_N(\hat{f}_{d,T}-p/N) \\ &+ \alpha_ID_M(\hat{f}_{B,I}-k/M) D_N(\hat{f}_{d,I}-p/N) + W(k,p) \nonumber
\end{align}
where $D_n(x) = \sin(n\pi x)/\sin(\pi x)$ is the Dirichlet function and $W(k,p)$ represents the 2-D FFT of noise.

One can observe from \eqref{Eq:RDmap} that the interference signal will form a sharp peak in the range-Doppler image. In particular, it is worth noting that, although the interference might be attributable to the transmission from the antenna sidelobe of one radar and receiving by the antenna sidelobe of the other, the potential interference level can be significantly higher than the target reflections due to the non-ideal antenna sidelobe characteristic.
This is attributable to the one-way propagation characteristic of the interference signal and the direct (without reflection) blast from one's transmission to the other's reception \cite{8461806}.
Specifically, according to the radar range-equation, the power of the target returns (\ie, $|\alpha_T|^2$) can be determined by
\begin{equation}\label{Eq:PowerTarget}
  P_{T,r} = \frac{P_tG_T^2 \lambda^2\sigma_t L_t}{(4\pi)^3 R_T^4},
\end{equation}
where $P_t$ is the system transmit power, $G_T$ is the antenna gain in the target direction, $\lambda = c/f_c$ is the wavelength, $c$ is the light speed,
$\sigma_t$ is the radar cross section (RCS) of the target, $L_t$ is the propagation loss, and $R_T$ denotes the target range. The power of interference (\ie, $|\alpha_I|^2$) is given by
\begin{equation}\label{Eq:PowerItf}
  P_{I,r} = \frac{P_tG_{t,I}G_{r,I} \lambda^2 L_I}{(4\pi)^2 R_I^2},
\end{equation}
where we have assumed the two automotive radar systems have the same transmit power, $G_{t,I}$ and $G_{r,I}$ denote the transmit and receive antenna gains associated with the interference, respectively, $L_I$ denotes the propagation loss, and $R_I$ is the range between the two radar systems. Thus, considering a target with RCS of $1 \textrm{m}^2$ and ignoring the propagation loss, we have the power ratio at the receiver input as,
\begin{equation}\label{Eq:PowerRatio}
  \frac{P_{I,r}}{P_{T,r}} = \frac{4\pi R_T^4 G_{t,I}G_{r,I}}{G_T^2 R_I^2}.
\end{equation}
Therefore, the non-negligible interference power will result in serious interference for both automotive radar systems.
%
%
In the following, we formulate the problem of mutual interference mitigation for a simple SISO scenario and then extend it to the MIMO case.

\section{SISO Mutual Interference Mitigation} \label{sec:siso}
\begin{figure}[!t]
\centering
\includegraphics[width = 0.6\textwidth,draft=false]{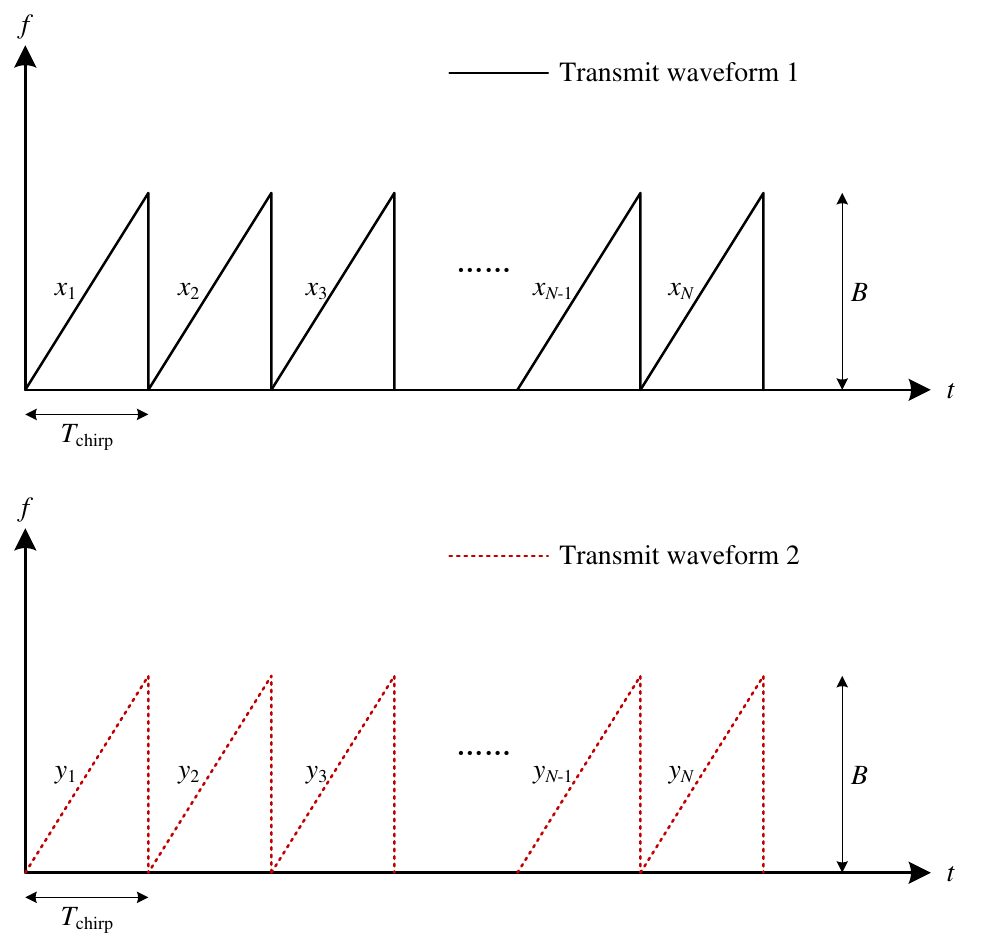}
\setlength{\abovecaptionskip}{0pt}
\caption{An illustration of the SISO coding scheme for two radar systems operating under same FMCW parameters.} \label{Fig:2}
\end{figure}

Consider two systems that use $N$ periodic consecutive pulses: $\mathbf{x}=[x_1, x_2. \cdots, x_N]^T$and $\mathbf{y}=[y_1, y_2. \cdots, y_N]^T$, respectively, as depicted in Fig.~\ref{Fig:2}.
That is to say, in the $n^{th}$ pulse of a CPI, the first radar system transmits $x_n u(t)$ and the second radar system transmits $y_n u(t)$. Moreover, to keep constant transmit power over the $N$ pulses, we constrain the code sequences to be unimodular, \ie, $|x_n| = |y_n| = 1, n=1,2, \cdots, N$.
With this coding scheme, the $m^{th}$ (fast-time) sample of the $n^{th}$ de-chirped signal is denoted by,
\begin{align}
    r^c(m,n) &= \alpha_T\exp(j2\pi(\hat{f}_{B,T}m+\hat{f}_{d,T}n)) \nonumber \\&+ \alpha_Ix_n^*y_{(n+l)\text{mod}~ N}\exp(j2\pi(\hat{f}_{B,I}m+\hat{f}_{d,I}n)) \nonumber \\&+ w(m,n).
\end{align}
for correlation lags $l \in \{-N+1, \cdots, N-1\}$.
The term $l$ has been introduced here in order to allow for asynchronous transmission of two radars.
The corresponding range-Doppler image, is thus given by
\begin{align}
    \text{RD}^c(k,p) = \alpha_TD_M(\hat{f}_{B,T}-k/M)
D_N(\hat{f}_{d,T}-p/N) \nonumber \\ + \alpha_ID_M(\hat{f}_{B,I}-k/M)
r_{xy}^l(\hat{f}_{d,I}-p/N) + W(k,p),
\end{align}
where
\begin{align}
    r_{xy}^l(f) = \sum_{n=1}^{N}{x_n^*y_{(n+l)\text{mod}~ N}\exp(j2\pi nf)}
\end{align}
is the periodic cross-ambiguity function (PCAF) of $\mathbf{x}$ and $\mathbf{y}$, and is to be minimized \cite{aubry2013ambiguity, chen2008mimo, he2011synthesizing}.

To suppress the interference power in the range-Doppler image, we aim at designing $\mbx$ and $\mby$ to minimize $r_{xy}^{l}(f)$ within a range of interest for $f$. To this end, we propose two methods to design $\mbx$ and $\mby$ in the following subsections.

\subsection{The Doppler-Shifting Scheme}
\begin{figure}[!t]
  \centering
  {\subfigure[]{{\includegraphics[width = 0.6\textwidth,draft=false]{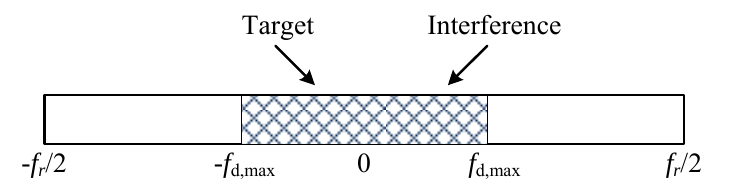}} \label{Fig:dopplershift_a}} }
  {\subfigure[]{{\includegraphics[width = 0.6\textwidth,draft=false]{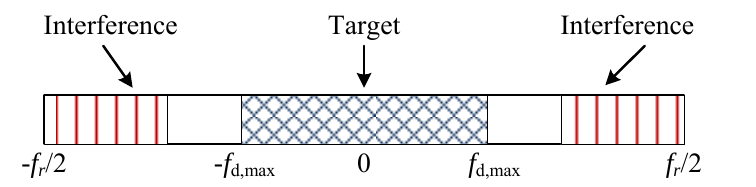}} \label{Fig:dopplershift_b}} }
  \caption{(a) The Doppler spectra of target and interference without coding and (b) the Doppler spectra of target and interference with Doppler-shifting coding.}\label{Fig:dopplershift}
\end{figure}

First, we propose a simple heuristic coding scheme to mitigate the interference. For simplicity and without loss of generality, we assume that the Doppler frequency of the interference signal satisfies $f_{D,I} \in [-f_{d,\max}, f_{d,\max}]$, where $f_{d,\max}$ denotes the maximum possible Doppler frequency (for both the target reflections and the interference signal). Note that for an automotive radar system with sweep time $T_c$, the unambiguous Doppler frequency that the system can identify is determined by
\begin{equation}
  \frac{-f_r}{2} \leq f_d \leq \frac{f_r}{2},
\end{equation}
with $f_r = {1}/{T_c}$ (we can treat $f_r$ as the pulse repetition frequency (PRF)). Assuming that $f_{d,\max} \leq {f_r}/{2}$,  the possibly occupied Doppler frequencies of the interference signal can be illustrated in Fig. \ref{Fig:dopplershift_a}. We can observe that, without slow-time coding, the Doppler frequency of target reflections and interference signal might occupy the same area and it results in mutual interference.

In order to mitigate the interference in the Doppler region, we introduce a coding scheme to shift the Doppler spectrum of the interference signal into the high frequency area ($>f_{d,\max}$ or $< -f_{d,\max}$). We call the resulting code the \emph{Doppler-shifting code}. With such a coding scheme, it is possible to  separate the target reflections and interference signal in the Doppler domain. As a result, we can apply low-pass filtering in the Doppler domain to mitigate the interference. To this end, we propose using the following codes to shift the Doppler spectrum of the interference signal:
\begin{align}\label{Eq:Coding1}
  \mbx &= [1,1,\cdots,1]^T,  \\
  \mby &=
  \begin{cases}
    [1,-1,\cdots,-1,1]^T, \textrm{if} \ N \ \textrm{is odd}, \\
    [1,-1,\cdots,1,-1]^T, \textrm{if} \ N \ \textrm{is even}.
  \end{cases}
\end{align}
Note that, the two sequences are orthogonal for even $N$ and quasi-orthogonal for odd $N$.
It is easy to verify that,
\begin{align}
  |r_{xy}^l| &= \left|\sum_{n=1}^N \exp(j n \pi ) \exp(j2\pi (\hat{f}_{d,I}-p/N_f)n)\right| \nonumber \\
        &= \left|\sum_{n=1}^N \exp(j2\pi ((\hat{f}_{d,I}+\frac{1}{2})-p/N_f)n)\right| \nonumber \\
        &= \left|\frac{\sin(N\pi (\hat{f}_{d,I}+1/2-p/N_f))}{\sin(\pi (\hat{f}_{d,I}+1/2-p/N_f))}\right|.
\end{align}
Therefore, the above codes enable the automotive radar to shift the Doppler frequency of interference signal from ${f}_{d,I}$ to ${f}_{d,I}+f_r/2$, as shown in Fig. \ref{Fig:dopplershift_b}. In particular, if $f_r$ satisfies
\begin{equation}\label{Eq:PRF}
  4f_{d,\max} < f_r,
\end{equation}
we can isolate the target reflections and the interference signal  in the Doppler domain with any Doppler frequency $f_{D,I} \in [-f_{d,\max}, f_{d,\max}]$.

On the other hand, note that the maximum value of the first sidelobe of the function $|\sin(N\pi f)/\sin(\pi f)|$ equals 
\begin{equation}\label{Eq:Sidelobe}
  \textrm{SLL}_1 = \frac{1}{\sin(\frac{3\pi}{2N})} \approx \frac{2N}{3\pi} (\textrm{for large}\ N).
\end{equation}
As a result, although the Doppler shifting coding scheme is simple, it may suffer from large sidelobes and the interference signal may not be extensively canceled.
Furthermore, the condition in \eqref{Eq:PRF} may constrain the radar system to identify only the slow-moving vehicles.

\subsection{Optimized Coding Scheme for SISO Radars} \label{subsec:opt}
In this subsection, we seek to optimize $\mbx$ and $\mby$ such that the corresponding $|r_{xy}^l(f)|$ has small values in a desired area. Given that the two radar systems usually have unsynchronized transmissions, the desired area should include all possible delays.
Hence, we consider the following optimization problem with respect to (w.r.t.) the two codes $\mathbf{x}$ and $\mathbf{y}$:
\begin{align}\label{eq:mainsiso}
    &\min_{\mathbf{x}, \mathbf{y}}{\sum_{l=-(N-1)}^{N-1} \sum_{p=-P}^{P} {|r_{lp}|^2}} \nonumber \\ &~ \text{s.t.} \qquad |x_n|=1,~ |y_n|=1,~ \forall\,n,
\end{align}
where
\begin{align}
    r_{lp} = \sum_{n=1}^{N}{x_n^*y_{(n+l)\text{mod}~ N}\exp(-j2\pi np/N_f)}
\end{align} is the discrete PCAF with
$N < N_f$,
$0 < P < N_f$,
$N_f$ is the overall number of discrete (Doppler) frequencies, and the value of $P$ is  closely related to the maximum Doppler frequency of interest and has to be chosen carefully so that multiple interferers can form peaks given they are all in the unambiguous region.
After some algebraic manipulation, the criterion in \eqref{eq:mainsiso} can be reformulated as
\begin{align}\label{eq:cross_amb_opt_periodic}
    &\min_{\mathbf{x}, \mathbf{y}}{\sum_{l=-(N-1)}^{N-1} \sum_{p=-P}^{P} {|\mathbf{x}^H \Diag{\mathbf{f}_p} \mathbf{C}_l \mathbf{y}|^2}} \nonumber \\ &~\text{s.t.} \qquad |x_n|=1,~ |y_n|=1,~ \forall\,n,
\end{align}
where $\mathbf{C}_l=\mathbf{C}_{-l}^T=\begin{bmatrix}\mathbf{0} & \mathbf{I}_{N-l} \\ \mathbf{I}_l & \mathbf{0}\end{bmatrix}$is a circular shift matrix and $n^{th}$ element of $\mathbf{f}_p$ is $\exp(-j2\pi np/N_f)$.

Note that the optimization problem in \eqref{eq:cross_amb_opt_periodic} is non-convex and appears to be difficult to solve \cite{Razaviyayn2013, 8454321}. Herein, we propose to tackle the problem in a cyclic manner. Specifically, in the $s^{th}$ iteration of our cyclic optimization process, we first optimize $\mbx$ for fixed $\mby^{(s-1)}$ and then optimize $\mby$ for fixed $\mbx^{(s)}$. In the following, we present the solution to the two sub-problems involved in each iteration of the cyclic approach. For notational simplicity, we omit the superscripts of $\mby^{(s-1)}$ and $\mbx^{(s)}$. 

\vspace{7pt}
$\bullet$ \textbf{Optimization of $\mbx$ for fixed $\mby$}:

The associated optimization problem can be recast as
\begin{align}\label{Eq:CrossAmbOptFixY_Periodic}
  \min_{\mbx} &~ \mbx^H \mbB_y \mbx,\nonumber\\
  \ \textrm{s.t.}&~ \ |x_n| = 1, ~~n = 1,2,\cdots,N,
\end{align}
where
\begin{equation}\label{Eq:By}
  \mbB_y = \sum_{l=-N+1}^{N-1}\sum_{p=-P}^{P} \Diag{\mbf_p} \mbC_l \mby \mby^H \mbC_l^H \Diag{\mbf_p}^H.
\end{equation}
The problem in \eqref{Eq:CrossAmbOptFixY_Periodic} is called a (non-convex) unimodular quadratic program (UQP). Such problems can be tackled by employing the \emph{power-method-like iterations} proposed in \cite{Soltanalian2013CREW} and \cite{Soltanalian2014UQP} (see also \cite{tang2015Relative,tang2016joint,naghsh2013unified} for more applications of  power-method-like iterations in radar code optimization).
Specifically, let $\gamma_y$ be a positive constant larger than the maximum eigenvalue of $\mathbf{B}_y$ to ensure $\mbD_y = \gamma_y\mbI_N - \mbB_y \succ \bzero$ (\ie, $\mbD_y$ is positive definite). It is easy to verify that the problem in \eqref{Eq:CrossAmbOptFixY_Periodic} can be equivalently written as:
\begin{align}\label{Eq:CrossAmbOptFixYrecast_Periodic2}
  \max_{\mbx}&\ \mbx^H \mbD_y \mbx, \nonumber\\\  \textrm{s.t.}& \ |x_n| = 1,~ n = 1,2,\cdots,N.
\end{align}

In the $t^{th}$ (inner) iteration, we update $\mbx$ by using the following power-method-like iterations (PMLI):
\begin{equation}\label{Eq:CrossAmbOptFixYSolution_Periodic}
  \mbx^{(s,t)} = \exp(j \arg(\mbD_y\mbx^{(s,t-1)})).
\end{equation}

\vspace{7pt}
$\bullet$ \textbf{Optimization of $\mby$ for fixed $\mbx$}:

The optimization of $\mby$ for fixed $\mbx$ is formulated as follows:
\begin{align}\label{Eq:CrossAmbOptFixXrecast_Periodic}
  \min_{\mby} &\ \mby^H \mbB_x \mby,\ \nonumber\\
  \textrm{s.t.} &\ |y_n| = 1, n = 1,2,\cdots,N,
\end{align}
where
\begin{equation}\label{Eq:Bx}
  \mbB_x = \sum_{l=-N+1}^{N-1}\sum_{p=-P}^{P} \mbC_l^H \Diag{\mbf_p}^H\mbx\mbx^H\Diag{\mbf_p}\mbC_l.
\end{equation}
Similar to the previous case, we can tackle the optimization problem in \eqref{Eq:CrossAmbOptFixXrecast_Periodic} iteratively. Specifically, the solution in the $t^{th}$ (inner) iteration is given by
\begin{equation}\label{Eq:CrossAmbOptFixXSolution_Periodic}
  \mby^{(s,t)} = \exp(j \arg(\mbD_x \mby^{(s,t-1)})),
\end{equation}
where $\mbD_x = \gamma_x\mbI_N - \mbB_x$ and $\gamma_x$ is a positive constant larger than the maximum eigenvalue of $\mathbf{B}_x$ to ensure $\mbD_x \succ \bzero$.

Finally, the steps of the proposed algorithm to minimize the discrete PCAF for two identical SISO systems is summarized in Algorithm \ref{Alg:1}.

\begin{remark}\textbf{(Optimality and the Convergence):}
The optimization problem in \eqref{eq:cross_amb_opt_periodic} is NP-hard and multimodal, \ie, the objective has multiple local optima \cite{7676417}. Due to the non-convex nature of the objective function, one usually settles for an approximation algorithm that yields local optima. In the proposed approach, we tackle the non-convexity of the problem by resorting to a cyclic minimization algorithm. In each half of the cycle, we optimize for one set of variables keeping the other fixed, and these two subproblems are solved based on a local optimization method, namely PMLI for UQP, that yields good local optima \cite{Soltanalian2014UQP}.
From \eqref{eq:cross_amb_opt_periodic}, it can be deduced that the objective value is lower bounded at $0$. Furthermore, from \eqref{Eq:CrossAmbOptFixY_Periodic} and \eqref{Eq:CrossAmbOptFixXrecast_Periodic}, we know that for $s$-th iteration \looseness =-1
\begin{align}
    \mbx^{(s+1)} &= \arg\min_{\mbx\in\Omega_x} ~ \mbx^H \mbB_y^{(s)} \mbx,~ \text{and}\\
    \mby^{(s+1)} &= \arg\min_{\mby\in\Omega_y} ~ \mby^H \mbB_x^{(s+1)} \mby.
\end{align}
where $\Omega_z = \{\mbz \mid |z_n|=1, \forall n \}$. Therefore, one can conclude that, in each iteration the objective value is monotonically decreasing, and that the algorithm converges. 
\end{remark}

\begin{remark}\textbf{(Computational Complexity):}
Note that, in the code optimization, one needs to calculate $\mbB_x$ and $\mbB_y$ at each iteration. Specifically, in the computation of $\mbB_y$, we need $N$ complex multiplications to obtain $\Diag{\mbf_p} \mbC_l \mby$ (whose  $n^{th}$ element is $e^{-j2\pi np/N_f}y_{(n+l)\mod N}$). As a result, the overall computational complexity of computing $\mbB_y$ is of the order $\Oh{N_f(2N-1)(N^2+N)}$. Similarly, the overall computational complexity of computing $\mbB_x$ is  of the order $\Oh{N_f(2N-1)(N^2+N)}$, which seems to be quite high. In Appendix \ref{app:apdA}, we provide a computationally fast and inexpensive way to calculate $\mbB_x$ and $\mbB_y$.
\end{remark}
In what follows, we extend our discussion to the more general case of MIMO automotive radar systems.

\begin{algorithm}[!t]
  \caption{ \small Automotive Radar Waveform Design Algorithm for Mutual Interference Mitigation (\emph{SISO Case})}\label{Alg:1}
  \begin{algorithmic}[1]
  \Init $\mbx^{(0)}$, $\mby^{(0)}$, $s=0$
  \Output\ $\mbx^{\star}$, $\mby^{\star}$
  \Repeat
    \State $s \gets s+1$
    \Statex \textbf{Update of $\mbx^{(s)}$:} 
    \State Calculate $\mbB_y^{(s)}$ with \eqref{Eq:By}
    \State $t = 0$, $\mbx^{(s,t)} = \mbx^{(s-1)}$
    \Repeat 
    \State $\mbx^{(s,t)} = \exp(j \arg((\gamma_y^{(s)}\mbI_N - \mbB_y^{(s)})\mbx^{(s,t-1)}))$
    \State $t \gets t+1$
    \Until{convergence} 
    \State $\mbx^{(s)} = \mbx^{(s,t)}$
    \Statex \textbf{Update of $\mby^{(s)}$:} 
    \State Calculate $\mbB_x^{(s)}$ with \eqref{Eq:Bx}
    \State $t = 0$, $\mby^{(s,t)} = \mby^{(s-1)}$
    \Repeat
    \State $\mby^{(s,t)} = \exp(j \arg((\gamma_x^{(s)}\mbI_N - \mbB_x^{(s)})\mby^{(s,t-1)}))$
    \State $t \gets t+1$
    \Until{convergence} 
    \State $\mby^{(s)} = \mby^{(s,t)}$
    \Until {a pre-defined stop criterion is satisfied, \eg, $|J^{(s)} - J^{(s-1)}| \leq \epsilon$, for some $\epsilon > 0$ where $J$ denotes the objective function of the problem \eqref{eq:cross_amb_opt_periodic}}
    \State $\mathbf{x}^{\star}=\mathbf{x}^{(s)}, \mathbf{y}^{\star}=\mathbf{y}^{(s)}$
    \end{algorithmic}
\end{algorithm}

\section{Extension to the  Generalized MIMO Scenario} \label{sec:mimo}
In a MIMO scenario, the mutual interference stems not only from the waveforms of a similar radar system nearby, but also from the various waveforms transmitted by the same radar system. In this case, our optimization problem for radar waveform design can be formulated as
\begin{align}\label{eq:mimooptmain}
    &\min_{\{\mathbf{x}_m\}, \{\mathbf{y}_k\}}\sum_{m,k}\sum_{l=-(N-1)}^{N-1} \sum_{p=-P}^{P} \left\lbrace |\mathbf{x}_m^H \Diag{\mathbf{f}_p} \mathbf{C}_l \mathbf{y}_k|^2 + \right.\nonumber \\
    &\qquad\qquad~\left.|\mathbf{x}_m^H \Diag{\mathbf{f}_p} \mathbf{C}_l \mathbf{x}_m|^2 + |\mathbf{y}_k^H \Diag{\mathbf{f}_p} \mathbf{C}_l \mathbf{y}_k|^2\right\rbrace \nonumber\\ &\quad~\text{s.t.} \qquad \mathbf{x}_m~\text{and}~\mathbf{y}_k~\text{are unimodular for all}~m,k,
\end{align}
in which $\{\mathbf{x}_m\}_{m=1}^{M}$ and $\{\mathbf{y}_k\}_{k=1}^{K}$ are the codes used for modulation on different antennas of the two radar systems. 
Note that the first term of \eqref{eq:mimooptmain} accounts for the mutual interference of different radar systems while the second and third terms account for the self-interference among the waveforms transmitted in the same radar systems.
Let 
\begin{align}\label{eq:Alp}
    \mathbf{A}_{l,p} &\triangleq \Diag{\mathbf{f}_p} \mathbf{C}_l, \\ 
    \mathbf{X}&=[\mathbf{x}_1, \cdots, \mathbf{x}_M], \\
    \mathbf{Y}&=[\mathbf{y}_1, \cdots, \mathbf{y}_K],
\end{align}
 and note that the objective in \eqref{eq:mimooptmain} can be written in a compact form as
\begin{align}\label{eq:Qmain}
    &Q(\mathbf{X}, \mathbf{Y}) \\&=\sum_{l,p} {\|\mathbf{X}^H\mathbf{A}_{l,p}\mathbf{X}\|_F^2 + \|\mathbf{Y}^H\mathbf{A}_{l,p}\mathbf{Y}\|_F^2 + \|\mathbf{X}^H\mathbf{A}_{l,p}\mathbf{Y}\|_F^2}.\nonumber
\end{align}
Tackling \eqref{eq:Qmain} appears to be more difficult than the optimization problem formulated in the SISO case in \eqref{eq:mainsiso}, as the new objective is \textit{quartic} in both radar codes ($\mathbf{X}$ and $\mathbf{Y}$) and the fact that the number of PCAF values to be suppressed is growing more quickly in terms of the problem dimension than the number of design variables. In the following, we formulate a quadratic alternative to \eqref{eq:mimooptmain} that can be tackled more efficiently.

\subsection{The Quartic to Quadratic Transformation}
In order to recast the problem in a quadratic form, let
\begin{align}\label{eq:14}
    \mathbf{A}_{l,p}^r &= \frac{1}{2}(\mathbf{A}_{l,p} + \mathbf{A}_{l,p}^H),\nonumber\\ \mathbf{A}_{l,p}^i &= \frac{1}{2}(\mathbf{A}_{l,p} - \mathbf{A}_{l,p}^H)
\end{align}
and note that
\begin{enumerate}
    \item Both matrices $\mathbf{A}_{l,p}^r$ and $j\mathbf{A}_{l,p}^i$ are Hermitian \cite{7676417}.
    \item For any generic vector $\mathbf{z}$,
    \begin{align} \label{eq:15}
        \mathbf{z}^H\mathbf{A}_{l,p}\mathbf{z} = \mathbf{z}^H\mathbf{A}_{l,p}^r\mathbf{z} + \mathbf{z}^H\mathbf{A}_{l,p}^i\mathbf{z}
    \end{align}
    where
    \begin{align} \label{eq:16}
        \mathbf{z}^H\mathbf{A}_{l,p}^r\mathbf{z} \in  \realR, \qquad
        j\mathbf{z}^H\mathbf{A}_{l,p}^i\mathbf{z} \in \realR.
    \end{align}
    In particular it follows from the above that
    \begin{align} \label{eq:17}
        |\mathbf{z}^H\mathbf{A}_{l,p}\mathbf{z}|^2 = |\mathbf{z}^H\mathbf{A}_{l,p}^r\mathbf{z}|^2 + |\mathbf{z}^Hj\mathbf{A}_{l,p}^i\mathbf{z}|^2.
    \end{align}
\end{enumerate}

We particularly observe that the quartic behavior of \eqref{eq:mimooptmain} and \eqref{eq:Qmain} stems from self-interference terms:
\begin{align}
    \{ |\mathbf{x}_m^H \mathbf{A}_{l,p} \mathbf{x}_m|^2 \}, \{ |\mathbf{y}_k^H \mathbf{A}_{l,p} \mathbf{y}_k|^2 \}
\end{align}
for all $m \in \{1, \cdots, M\}$ and $k \in \{1, \cdots, K\}$.

Based on \eqref{eq:14}-\eqref{eq:17}, one can write
\begin{align}\label{eq:39}
    \sum_{l,p}|\mathbf{x}_m^H \mathbf{A}_{l,p} \mathbf{x}_m|^2 &= \sum_{l,p}|\mathbf{x}_m^H \mathbf{A}_{l,p}^r \mathbf{x}_m|^2 + |\mathbf{x}_m^H j\mathbf{A}_{l,p}^i \mathbf{x}_m|^2 \nonumber\\
    &= \sum_{l,p}|\mathbf{x}_m^H (\mathbf{A}_{l,p}^r + \zeta\mbI_N) \mathbf{x}_m - \zeta N|^2 \nonumber \\ &\qquad+ |\mathbf{x}_m^H (j\mathbf{A}_{l,p}^i + \zeta\mbI_N) \mathbf{x}_m - \zeta N|^2 \nonumber \\
    &= \sum_{l,p}|\mathbf{x}_m^H \mathbf{\tilde A}_{l,p}^r \mathbf{x}_m - \zeta N|^2 \nonumber \\ &\qquad+ |\mathbf{x}_m^H \mathbf{\tilde A}_{l,p}^i \mathbf{x}_m - \zeta N|^2,
\end{align}
where 
\begin{align}\label{eq:tildea}
    \mathbf{\tilde A}_{l,p}^r &= \mathbf{A}_{l,p}^r + \zeta\mbI_N, \nonumber\\ \mathbf{\tilde A}_{l,p}^i &= j\mathbf{A}_{l,p}^i + \zeta\mbI_N,    
\end{align}
and $\zeta \in \realR$ is chosen such that
\begin{align}\label{eq:zeta}
    \zeta > - \min \left( \bigcup_{l,p}\left\lbrace \gamma_{\min}\left(\mathbf{A}_{l,p}^r\right), \gamma_{\min}\left(j\mathbf{A}_{l,p}^i\right) \right\rbrace\right)
\end{align}
to ensure the positive definiteness of $\{\mathbf{\tilde A}_{l,p}^r\}$ and $\{\mathbf{\tilde A}_{l,p}^i\}$, where $\gamma_{\min}(\cdot)$ denotes the minimum eigenvalue of its matrix argument.
Observe that the quantity in \eqref{eq:39} is still quartic w.r.t. $\mathbf{x}_m$, which in fact is difficult to minimize.
A quadratic alternative, however, can be proposed in the following manner.
Note that the quantity in \eqref{eq:39} will be made small when the quadratic quantities $\{\mathbf{x}_m^H \mathbf{\tilde A}_{l,p}^r \mathbf{x}_m\}$ and $\{\mathbf{x}_m^H \mathbf{\tilde A}_{l,p}^i \mathbf{x}_m\}$ are  \textit{close} to $\zeta N$.
This is only possible when unit-norm vectors $\{\mathbf{u}_{l,p,m}^r\}$ and $\{\mathbf{u}_{l,p,m}^i\}$ exist such that $(\mathbf{\tilde A}_{l,p}^r)^{1/2}\mathbf{x}_m$ is \textit{close} to $\sqrt{\zeta N}\mathbf{u}_{l,p,m}^r$, and likewise $(\mathbf{\tilde A}_{l,p}^i)^{1/2}\mathbf{x}_m$ is \textit{close} to $\sqrt{\zeta N}\mathbf{u}_{l,p,m}^i$. 
As a result, minimization of \eqref{eq:39} can be approached by a reformulation, in the form of the following alternative quadratic optimization problem \cite{5072243, 1377501, 4567663, SOLTANALIAN2014132}:
\begin{align}\label{eq:optxpart}
    \min  ~ & \sum_{l,p} \left\lbrace \left\|(\mathbf{\tilde A}_{l,p}^r)^{1/2}\mathbf{x}_m - \sqrt{\zeta N} \mathbf{u}_{l,p,m}^r\right\|_2^2 \right.\nonumber\\
    & \qquad\qquad+\left.\left\|(\mathbf{\tilde A}_{l,p}^i)^{1/2}\mathbf{x}_m - \sqrt{\zeta N}\mathbf{u}_{l,p,m}^i \right\|_2^2 \right\rbrace\nonumber\\
     \st~&\mathbf{x}_m~\text{are unimodular for all}~m, \nonumber \\
    & \|\mathbf{u}_{l,p,m}^r\|_2=\|\mathbf{u}_{l,p,m}^i\|_2 = 1  ~\text{for all}~l,p,m.
\end{align}
Thus the criteria in \eqref{eq:39} and \eqref{eq:optxpart} are ``almost equivalent'' in the sense that their minimization is likely to lead to signals with similar properties in terms of mutual interference.
Interestingly, one can observe that \eqref{eq:optxpart} is quadratic instead of quartic--- a transformation that was made possible by judicious over-parametrization. In a similar manner, we argue that $\{|\mathbf{y}_k^H \mathbf{A}_{l,p} \mathbf{y}_k|^2\}$ can be made small by solving the alternative problem:
\begin{align}\label{eq:optypart}
    \min ~ & \sum_{l,p}  \left\lbrace \left\|(\mathbf{\tilde A}_{l,p}^r)^{1/2}\mathbf{y}_k - \sqrt{\zeta N} \mathbf{v}_{l,p,k}^r\right\|_2^2 \right.\nonumber\\
    & \qquad\qquad+\left.\left\|(\mathbf{\tilde A}_{l,p}^i)^{1/2}\mathbf{y}_k - \sqrt{\zeta N}\mathbf{v}_{l,p,k}^i \right\|_2^2 \right\rbrace\nonumber\\
    \st~&\mathbf{y}_k~\text{are unimodular for all}~k, \nonumber \\
    & \|\mathbf{v}_{l,p,k}^r\|_2=\|\mathbf{v}_{l,p,k}^i\|_2 = 1  ~\text{for all}~l,p,k.
\end{align}

As a result, the objective in \eqref{eq:mimooptmain} can be recast in its almost-equivalent quadratic form as described in \eqref{eq:mainopt} which is shown on the top of the next page.
\begin{figure*}[!htbp]
\begin{align}\label{eq:mainopt}
    \min_{\substack{\{\mathbf{x}_m\}, \{\mathbf{y}_k\},\\ \{\mathbf{u}_{l,p,m}^r\}, \{\mathbf{u}_{l,p,m}^i\}, \\ \{\mathbf{v}_{l,p,k}^r\}, \{\mathbf{v}_{l,p,k}^i\}}}~&\sum_{l,p}\sum_{m\neq k} \{|\mathbf{x}_m^H \mathbf{A}_{l,p} \mathbf{x}_k|^2 + |\mathbf{y}_m^H \mathbf{A}_{l,p} \mathbf{y}_k|^2\} + \sum_{l,p}\sum_{m,k} \{|\mathbf{x}_m^H \mathbf{A}_{l,p} \mathbf{y}_k|^2\} \nonumber \\
    &\qquad+~\sum_{l,p}\left[ \sum_{m}\left\lbrace\left\|(\mathbf{\tilde A}_{l,p}^r)^{1/2}\mathbf{x}_m - \sqrt{\zeta N} \mathbf{u}_{l,p,m}^r \right\|_2^2 + \left\|(\mathbf{\tilde A}_{l,p}^i)^{1/2}\mathbf{x}_m - \sqrt{\zeta N}\mathbf{u}_{l,p,m}^i\right\|_2^2 \right\rbrace \right.\nonumber\\
    &\qquad\qquad + \left.\sum_{k}\left\lbrace \left\|(\mathbf{\tilde A}_{l,p}^r)^{1/2}\mathbf{y}_k - \sqrt{\zeta N} \mathbf{v}_{l,p,k}^r\right\|_2^2 + \left\|(\mathbf{\tilde A}_{l,p}^i)^{1/2}\mathbf{y}_k - \sqrt{\zeta N}\mathbf{v}_{l,p,k}^i\right\|_2^2 \right\rbrace \right] \nonumber\\
    \text{s.t.}~~~~&\mathbf{x}_m~\text{and}~\mathbf{y}_k~\text{are unimodular for all}~m,k, \nonumber \\
    & \|\mathbf{u}_{l,p,m}^r\|_2=\|\mathbf{u}_{l,p,m}^i\|_2 = 1  ~\text{for all}~l,p,m, \nonumber \\
    & \|\mathbf{v}_{l,p,k}^r\|_2~=\|\mathbf{v}_{l,p,k}^i\|_2 ~= 1~\text{for all}~l,p,k.
\end{align}
\hrule
\end{figure*}
Note that the optimization problem in \eqref{eq:mainopt} is still non-convex, especially because of the unimodular constraints imposed on $\{\mathbf{x}_m\}$ and $\{\mathbf{y}_k\}$. In the following subsection, we provide an efficient way to tackle the above problem for all individual optimization variables.

\subsection{The Optimization Procedure}
In order to efficiently tackle the problem in \eqref{eq:mainopt}, we resort to a cyclic optimization framework. Namely, we iteratively optimize the criterion with respect to one of the variables while keeping the rest of them fixed. In $s^{th}$ iteration, we separate each variable $\{\mathbf{x}_m\}$, $\{\mathbf{y}_k\}$, $\{\mathbf{u}_{l,p,m}^c\}, \{\mathbf{v}_{l,p,k}^c\}$ for all $m \in\{1, \cdots, M\}, k\in\{1, \cdots, K\}, l \in\{-(N-1), \cdots, N-1\}, p \in\{-P, \cdots, P\}, c \in \{r,i\}$ from the objective function in \eqref{eq:mainopt} and optimize them individually while fixing all other variables to their values from $(s-1)^{th}$ iteration.
In the following subsections, we describe such a process of variable separation and the corresponding solution techniques.
We drop the superscript $(s)$ for notational simplicity.

\vspace{5pt}
 $\bullet$ \textbf{Optimization of $\{\mathbf{x}_m\}_{m=1}^{M}$}:
\vspace{5pt}

We begin by reformulating the optimization problem in \eqref{eq:mainopt} w.r.t.  each of  $\mathbf{x}_m$ for all $m$. Eliminating all the other variables that do not depend on $\mathbf{x}_m$, the objective function in \eqref{eq:mainopt} becomes what is described in \eqref{eq:fulloptxm},  shown on the top of the next page,
\begin{figure*}[!hpbt]
\begin{align}\label{eq:fulloptxm}
    Q_{\mathbf{x}_m}    &= \underbrace{\mathbf{x}_m^H \left(\sum_{m'\neq m}\sum_{l,p}\mathbf{A}_{l,p}\mathbf{x}_{m'}\mathbf{x}_{m'}^H\mathbf{A}_{l,p}^H \right) \mathbf{x}_m}_{M-1~\text{terms}} + \underbrace{\mathbf{x}_m^H \left(\sum_{k}\sum_{l,p}\mathbf{A}_{l,p}\mathbf{y}_k\mathbf{y}_k^H\mathbf{A}_{l,p}^H \right) \mathbf{x}_m}_{K~\text{terms}} + \mathbf{x}_m^H \left(\sum_{l,p}\mathbf{\tilde A}_{l,p}^r +\mathbf{\tilde A}_{l,p}^i \right) \mathbf{x}_m \nonumber \\
    &\qquad - 2\sqrt{\zeta N}\Re\left\lbrace\mathbf{x}_m^H \sum_{l,p} (\mathbf{\tilde A}_{l,p}^r)^{H/2}\mathbf{u}_{l,p,m}^{r}\right\rbrace - 2\sqrt{\zeta N}\Re\left\lbrace\mathbf{x}_m^H \sum_{l,p} (\mathbf{\tilde A}_{l,p}^i)^{H/2}\mathbf{u}_{l,p,m}^{i}\right\rbrace + \text{const.}
\end{align}
\hrule
\end{figure*}
 or simply,
\begin{align}\label{eq:optxm}
    Q_{\mathbf{x}_m} = \mathbf{x}_m^H \mathbf{R}_{\mathbf{x}_m} \mathbf{x}_m + 2\Re\{\mathbf{x}_m^H \mathbf{s}_{\mathbf{x}_m}\} + \text{const.},
\end{align}
where
\begin{align}\label{eq:Rxm}
    \mathbf{R}_{\mathbf{x}_m} &= \sum_{m'\neq m}\sum_{l,p}\mathbf{A}_{l,p}\mathbf{x}_{m'}\mathbf{x}_{m'}^H\mathbf{A}_{l,p}^H + \sum_{k}\sum_{l,p}\mathbf{A}_{l,p}\mathbf{y}_k\mathbf{y}_k^H\mathbf{A}_{l,p}^H \nonumber\\ &\qquad\qquad+ \sum_{l,p}\mathbf{\tilde A}_{l,p}^r + \mathbf{\tilde A}_{l,p}^i
\end{align}
and
\begin{align}\label{eq:sxm}
    &\mathbf{s}_{\mathbf{x}_m} =\\ &~~ - \sqrt{\zeta N} \sum_{l,p}\left((\mathbf{\tilde A}_{l,p}^r)^{H/2}\mathbf{u}_{l,p,m}^{r} + (\mathbf{\tilde A}_{l,p}^i)^{H/2}\mathbf{u}_{l,p,m}^{i}\right). \nonumber
\end{align}
By dropping the constant term in \eqref{eq:optxm}, the objective function can be reformulated as,
\begin{align}
    Q_{\mathbf{x}_m} &= \mathbf{x}_m^H \mathbf{R}_{\mathbf{x}_m} \mathbf{x}_m + 2\Re\{\mathbf{x}_m^H \mathbf{s}_{\mathbf{x}_m}\} \nonumber\\
    &=\begin{bmatrix} \mathbf{x}_m \\ 1\end{bmatrix}^H \begin{bmatrix} \mathbf{R}_{\mathbf{x}_m} & \mathbf{s}_{\mathbf{x}_m} \\ \mathbf{s}_{\mathbf{x}_m}^H & 0\end{bmatrix} \begin{bmatrix} \mathbf{x}_m \\ 1 \end{bmatrix} \nonumber\\ 
    &= \bar{\mathbf{x}}_m^H \mathbf{B}_{\mathbf{x}_m} \bar{\mathbf{x}}_m
\end{align}
where 
\begin{align}
    \bar{\mathbf{x}}_m &\triangleq [\mathbf{x}_m~ 1]^T, \label{eq:xbarm}\\
    \mathbf{B}_{\mathbf{x}_m} &\triangleq \begin{bmatrix} \mathbf{R}_{\mathbf{x}_m} & \mathbf{s}_{\mathbf{x}_m} \\ \mathbf{s}_{\mathbf{x}_m}^H & 0\end{bmatrix}. \label{eq:Bxm}
\end{align}
Hence, the minimization of \eqref{eq:mainopt} w.r.t. $\mathbf{x}_m$ is equivalent to the following,
\begin{align} \label{eq:optxmshort}
    &\min_{\bar{\mathbf{x}}_m} ~~\bar{\mathbf{x}}_m^H \mathbf{B}_{\mathbf{x}_m} \bar{\mathbf{x}}_m \nonumber\\ &~\text{s.t.}~~|x_n(m)|=1, ~~n=1,\cdots,N, \nonumber\\ &\qquad~ \bar{\mathbf{x}}_m = \begin{bmatrix} \mathbf{x}_m \\ 1 \end{bmatrix}.
\end{align}
As a result of the unimodular constraint on $\mathbf{x}_m$, the term $\bar{\mathbf{x}}_m$ also has a constant $\ell_2$-norm, and hence, a diagonal loading of $\mathbf{B}_{\mathbf{x}_m}$ will not change the solution to the above problem \cite{7676417}. 
Therefore, \eqref{eq:optxmshort} can be rewritten in the following equivalent form:
\begin{align}
    &\max_{\bar{\mathbf{x}}_m} ~~\bar{\mathbf{x}}_m^H \mathbf{D}_{\mathbf{x}_m} \bar{\mathbf{x}}_m \nonumber\\ &~\text{s.t.}~~|x_n(m)|=1, ~~n=1,\cdots,N, \nonumber\\ &\qquad~\bar{\mathbf{x}}_m = \begin{bmatrix} \mathbf{x}_m \\ 1 \end{bmatrix},
\end{align}
where 
\begin{align}\label{eq:Dxm}
    \mathbf{D}_{\mathbf{x}_m} \triangleq \gamma_{\mathbf{x}_m} \mathbf{I}_{(N+1)}-\mathbf{B}_{\mathbf{x}_m},
\end{align}
with $\gamma_{\mathbf{x}_m}$ being larger than the maximum eigenvalue of $\mathbf{B}_{\mathbf{x}_m}$.
Note that the above problem similarly belongs to the family of UQPs \cite{Soltanalian2014UQP}, and can be efficiently tackled in an iterative manner using power-method-like iterations of the form \cite{7676417}:
\begin{align}\label{eq:xm_update}
    \mathbf{x}_m^{(s,t)} = \exp\left\lbrace j\arg \left( \begin{bmatrix} \mathbf{I}_{N\times N} \\ \mathbf{0}_{1\times N} \end{bmatrix}^T \mathbf{D}_{\mathbf{x}_m} \bar{\mathbf{x}}_m^{(s,t-1)}\right)\right\rbrace,
\end{align}
where the iterations can be initialized with the latest design of $\mathbf{x}_m$ (used as $\mathbf{x}_m^{(s,0)}$) and $t$ denotes the inner iteration number.

\vspace{7pt}
$\bullet$ \textbf{Optimization of $\{\mathbf{y}_k\}_{k=1}^{K}$}:
\vspace{5pt}

In order to solve \eqref{eq:mainopt} for $\mathbf{y}_k$ for all $k$, we follow the same algebraic manipulation with slight modifications.
In this case, the objective function, $Q_{\mathbf{y}_k}$, becomes 
\begin{align}\label{eq:optyk}
    Q_{\mathbf{y}_k} = \mathbf{y}_k^H \mathbf{R}_{\mathbf{y}_k} \mathbf{y}_k + 2\Re\{\mathbf{y}_k^H \mathbf{s}_{\mathbf{y}_k}\} + \text{const.}
\end{align}
where
\begin{align}\label{eq:Ryk}
    \mathbf{R}_{\mathbf{y}_k} &= \sum_{k'\neq k}\sum_{l,p}\mathbf{A}_{l,p}\mathbf{y}_{k'}\mathbf{y}_{k'}^H\mathbf{A}_{l,p}^H + \sum_{m}\sum_{l,p}\mathbf{A}_{l,p}^H\mathbf{x}_m\mathbf{x}_m^H\mathbf{A}_{l,p} \nonumber\\ &\qquad\qquad+ \sum_{l,p}\mathbf{\tilde A}_{l,p}^r + \mathbf{\tilde A}_{l,p}^i
\end{align}
and
\begin{align}\label{eq:syk}
    &\mathbf{s}_{\mathbf{y}_k} =\\ &~~~ - \sqrt{\zeta N} \sum_{l,p}\left((\mathbf{\tilde A}_{l,p}^r)^{H/2}\mathbf{v}_{l,p,k}^{r} + (\mathbf{\tilde A}_{l,p}^i)^{H/2}\mathbf{v}_{l,p,k}^{i}\right). \nonumber
\end{align}
As a result, we can  formulate a UQP for each $\mathbf{y}_k$ in a similar manner. The corresponding solution can be approached iteratively using the power-method-like recursions of the form
\begin{align}\label{eq:yk_update}
    \mathbf{y}_k^{(s,t)} = \exp\left\lbrace j\arg \left( \begin{bmatrix} \mathbf{I}_{N\times N} \\ \mathbf{0}_{1\times N} \end{bmatrix}^T \mathbf{D}_{\mathbf{y}_k} \bar{\mathbf{y}}_k^{(s,t-1)}\right)\right\rbrace,
\end{align}
where 
\begin{align}\label{eq:Dyk}
    \mathbf{D}_{\mathbf{y}_k} \triangleq \gamma_{\mathbf{y}_k} \mathbf{I}_{(N+1)}-\mathbf{B}_{\mathbf{y}_k}
\end{align}
with $\gamma_{\mathbf{y}_k}$ being larger than the maximum eigenvalue of $\mathbf{B}_{\mathbf{y}_k}$, and
\begin{align}
    \bar{\mathbf{y}}_k &\triangleq [\mathbf{y}_k ~ 1 ]^T \label{eq:ybark}\\
    \mathbf{B}_{\mathbf{y}_k} &\triangleq \begin{bmatrix} \mathbf{R}_{\mathbf{y}_k} & \mathbf{s}_{\mathbf{y}_k} \\ \mathbf{s}_{\mathbf{y}_k}^H & 0\end{bmatrix}. \label{eq:Byk}
\end{align}

$\bullet$ \textbf{Optimization of $\{\mathbf{u}_{l,p,m}^c\}$ and  $\{\mathbf{v}_{l,p,k}^c\}$}:
\vspace{5pt}

Solving \eqref{eq:mainopt} w.r.t. $\{\mathbf{u}_{l,p,m}^c\}$ and $\{\mathbf{v}_{l,p,k}^c\}$ for $c \in \{r,i\}$ is immediate and resolves into closed-form solution as follows:
\begin{align}
    \widehat{\mathbf{u}}_{l,p,m}^r &= \frac{(\mathbf{\tilde A}_{l,p}^r)^{1/2}\mathbf{x}_m}{\|(\mathbf{\tilde A}_{l,p}^r)^{1/2}\mathbf{x}_m\|_2},
    \label{eq:ufirst}\\
    \widehat{\mathbf{u}}_{l,p,m}^i &= \frac{(\mathbf{\tilde A}_{l,p}^i)^{1/2}\mathbf{x}_m}{\|(\mathbf{\tilde A}_{l,p}^i)^{1/2}\mathbf{x}_m\|_2},
    \\
    \widehat{\mathbf{v}}_{l,p,k}^r &= \frac{(\mathbf{\tilde A}_{l,p}^r)^{1/2}\mathbf{y}_k}{\|(\mathbf{\tilde A}_{l,p}^r)^{1/2}\mathbf{y}_k\|_2},
    \\
    \widehat{\mathbf{v}}_{l,p,k}^i &= \frac{(\mathbf{\tilde A}_{l,p}^i)^{1/2}\mathbf{y}_k}{\|(\mathbf{\tilde A}_{l,p}^i)^{1/2}\mathbf{y}_k\|_2},
    \label{eq:vlast}
\end{align}
for all $m \in\{1, \cdots, M\}, k\in\{1, \cdots, K\}, l \in\{-(N-1), \cdots, N-1\}, p \in\{-P, \cdots, P\}$.

\vspace{5pt}

Finally, the steps of the proposed algorithm for interference mitigation in the MIMO setting is summarized in Algorithm~\ref{Alg:2}.

\begin{remark} \textbf{(Convergence):}
As mentioned earlier for Algorithm \ref{Alg:1}, the Algorithm \ref{Alg:2} as well resorts to a cyclic optimization method to tackle the non-convexity of the problem \eqref{eq:mainopt}. For each iteration of Algorithm \ref{Alg:2}, one can observe that the objective value is monotonically decreasing and bounded from below, leading to the convergence of the algorithm.
Note that the final output of the cyclic algorithms often depends on the initialization of all the optimization variables.
Different initial points in the search space may lead to different final designs due to the non-convexity of the landscape.
For this reason, it is desirable to run Algorithm \ref{Alg:2} multiple times. A good candidate can be using the output of the current design as the initialization for the next design.
\end{remark}

\begin{remark}\textbf{(Computational Complexity and Parallelization):}
Note that, in the step 3 of the Algorithm \ref{Alg:2}, the calculation of $\tilde{\mathbf{A}}$ requires $\Oh{N_f(2N-1)(N + N^3)}$ number of complex multiplications. The term $N^3$ comes from the eigenvalue decomposition of the $N\times N$ matrix $\mathbf{A}$. Furthermore, the overall computational complexity of calculating $\mathbf{D}_{\mathbf{x}_m}$ and $\mathbf{D}_{\mathbf{y}_k}$ is $\Oh{(M+K)N_f(2N-1)[(M+K-1)N+N^3]}$.
However, it is interesting to note that the computation of $\tilde{\mathbf{A}}$ is required only once in the entire optimization procedure and can be performed in parallel. 
Moreover, in the $s^{th}$ iteration of the algorithm, solving for $\{\mathbf{u}^{c(s)}_{l,p,m}\}$, $\{\mathbf{v}^{c(s)}_{l,p,k}\}$ can also be done in parallel making the algorithm significantly more efficient from a computational viewpoint.
\end{remark}

It is interesting to note that the SISO objective considered in \eqref{eq:cross_amb_opt_periodic} is not a special case of the MIMO counterpart in \eqref{eq:mimooptmain} as the objective in \eqref{eq:cross_amb_opt_periodic} does not take the self-interference terms into account.
However, \eqref{eq:mimooptmain} can be reduced to a SISO scenario assuming $M=K=1$ as stated in the following:
\begin{align}\label{eq:sisooptmainalt}
    &\min_{\mathbf{x}, \mathbf{y}}\sum_{l=-(N-1)}^{N-1} \sum_{p=-P}^{P} \left\lbrace |\mathbf{x}^H \Diag{\mathbf{f}_p} \mathbf{C}_l \mathbf{y}|^2 + \right.\nonumber \\
    &\qquad\qquad~\left.|\mathbf{x}^H \Diag{\mathbf{f}_p} \mathbf{C}_l \mathbf{x}|^2 + |\mathbf{y}^H \Diag{\mathbf{f}_p} \mathbf{C}_l \mathbf{y}|^2\right\rbrace \nonumber\\ &\quad~\text{s.t.} \qquad \mathbf{x}~\text{and}~\mathbf{y}~\text{are unimodular,}
\end{align}
where both the self-interference and mutual interference terms are included in the objective,
and can be minimized efficiently using Algorithm \ref{Alg:2}.
In the next section, several numerical examples are provided to showcase the performance and efficiency of the proposed waveform design schemes.

\begin{algorithm}[!t]
  \caption{ \small Automotive Radar Waveform Design Algorithm for Interference Mitigation (\emph{MIMO Case})}\label{Alg:2}
  \begin{algorithmic}[1]
    \Init $\{\mbx_m^{(0)}\}$, $\{\mby_k^{(0)}\}$,  $\{\mathbf{u}_{l,p,m}^{r(0)}\}$,  $\{\mathbf{u}_{l,p,m}^{i(0)}\}$, $\{\mathbf{v}_{l,p,k}^{r(0)}\}$,  $\{\mathbf{v}_{l,p,k}^{i(0)}\}$, 
    for $m \in\{1, \cdots, M\}, k\in\{1, \cdots, K\}, l \in\{-(N-1), \cdots, N-1\}, p \in\{-P, \cdots, P\}$, $s=0$
    \Output $\{\mbx_m^{\star}\}_{m=1}^M, \{\mby_k^{\star}\}_{k=1}^K$
    \Repeat
    \State $s \gets s+1$
    \State Calculate $\mathbf{\tilde A}_{l,p}^{r(s)}, \mathbf{\tilde A}_{l,p}^{i(s)}$ for all $l, p$ following \eqref{eq:Alp},  \eqref{eq:14}, and \eqref{eq:tildea}
    \State Calculate $\zeta^{(s)}$ using \eqref{eq:zeta}
    
    \Statex \textbf{Update of $\{\mathbf{x}_m^{(s)}\}_{m=1}^{M}$:}
    \For{$m=1$ \To $M$}
        \State Calculate $\mathbf{R}_{\mathbf{x}_m}^{(s)}, \mathbf{s}_{\mathbf{x}_m}^{(s)}$ using \eqref{eq:Rxm} and \eqref{eq:sxm}
        \State Calculate $\bar{\mathbf{x}}_m^{(s)}, \mathbf{B}_{\mathbf{x}_m}^{(s)}$ using \eqref{eq:xbarm} and \eqref{eq:Bxm}
        \State Calculate $\mathbf{D}_{\mathbf{x}_m}^{(s)}$ using \eqref{eq:Dxm}
        \State $t = 0$, $\mbx_m^{(s,t)} = \mbx_m^{(s-1)}$
        \Repeat
            \State Calculate $\mathbf{x}_m^{(s,t)}$ using \eqref{eq:xm_update}
            \State $t \gets t+1$
        \Until{convergence}
        \State $\mbx_m^{(s)} = \mbx_m^{(s,t)}$
    \EndFor
    
    \Statex \textbf{Update of $\{\mathbf{y}_k^{(s)}\}_{k=1}^{K}$:}
    \For{$k=1$ \To $K$}
        \State Calculate $\mathbf{R}_{\mathbf{y}_k}^{(s)}, \mathbf{s}_{\mathbf{y}_k}^{(s)}$ using \eqref{eq:Ryk} and \eqref{eq:syk}
        \State Calculate $\bar{\mathbf{y}}_k^{(s)}, \mathbf{B}_{\mathbf{y}_k}^{(s)}$ using \eqref{eq:ybark} and \eqref{eq:Byk}
        \State Calculate $\mathbf{D}_{\mathbf{y}_k}^{(s)}$ using \eqref{eq:Dyk}
        \State $t = 0$, $\mby_k^{(s,t)} = \mby_k^{(s-1)}$
        \Repeat
            \State Calculate $\mathbf{y}_k^{(s,t)}$ using \eqref{eq:yk_update}
            \State $t \gets t+1$
        \Until{convergence}
        \State $\mby_k^{(s)} = \mby_k^{(s,t)}$
    \EndFor
    
    \Statex \textbf{Update of $\{\mathbf{u}_{l,p,m}^{c(s)}\}$ and  $\{\mathbf{v}_{l,p,k}^{c(s)}\}$ for all $c \in\{r, i\}$:}
    \State Calculate $\mathbf{u}_{l,p,m}^{r(s)}, \mathbf{u}_{l,p,m}^{i(s)}, \mathbf{v}_{l,p,k}^{r(s)}, \mathbf{v}_{l,p,k}^{i(s)}$ for each $m \in\{1, \cdots, M\}, k\in\{1, \cdots, K\}, l \in\{-(N-1), \cdots, N-1\}, p \in\{-P, \cdots, P\}$ using \eqref{eq:ufirst}-\eqref{eq:vlast}
    
    \Until {a pre-defined stop criterion is satisfied, \eg, $|\bar{J}^{(s)} - \bar{J}^{(s-1)}| \leq \epsilon$, for some $\epsilon > 0$ where $\bar{J}$ denotes the objective function of the problem \eqref{eq:mainopt}}
    \State $\{\mbx_m^{\star}\}_{m=1}^M = \{\mathbf{x}_m^{(s)}\}_{m=1}^M, \{\mby_k^{\star}\}_{k=1}^K = \{\mathbf{y}_k^{(s)}\}_{k=1}^K$
    \end{algorithmic}
\end{algorithm}


\section{Numerical Examples} \label{sec:num}
In the following, we begin with demonstrating the effectiveness of the two coding schemes described in Section \ref{sec:siso} for the SISO scenario using several examples. 
We then provide a similar numerical analysis for the MIMO case detailed in Section \ref{sec:mimo}.

\subsection{The SISO Scenario}
Consider two identical FMCW radar systems with the same carrier frequency of $f_\textrm{c} = 24$ GHz.
The bandwidth of the chirp signal is $B = 150$ MHz. The sweep time is $T_c =50 \ \mu$s.
The number of periods within a CPI is $N=256$.
we initialize our algorithm with normally distributed randomly generated codes (for $\mbx$ and $\mby$, respectively,) in the optimized coding scheme. 
\begin{figure}[!t]
  \centering
  {\subfigure[]{{\includegraphics[width = 0.35\textwidth,draft=false]{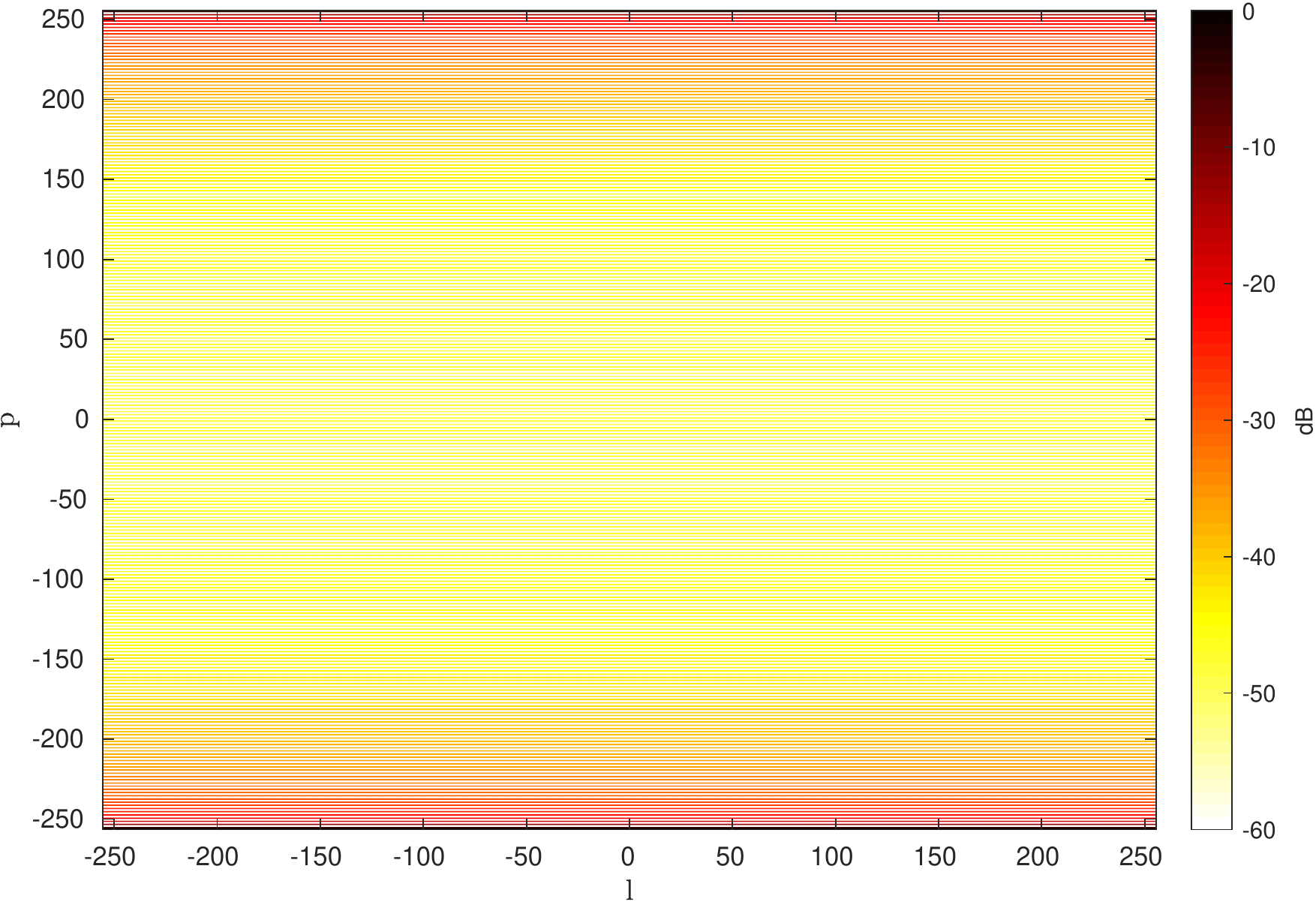}} \label{Fig:3a}} }
  {\subfigure[]{{\includegraphics[width = 0.35\textwidth,draft=false]{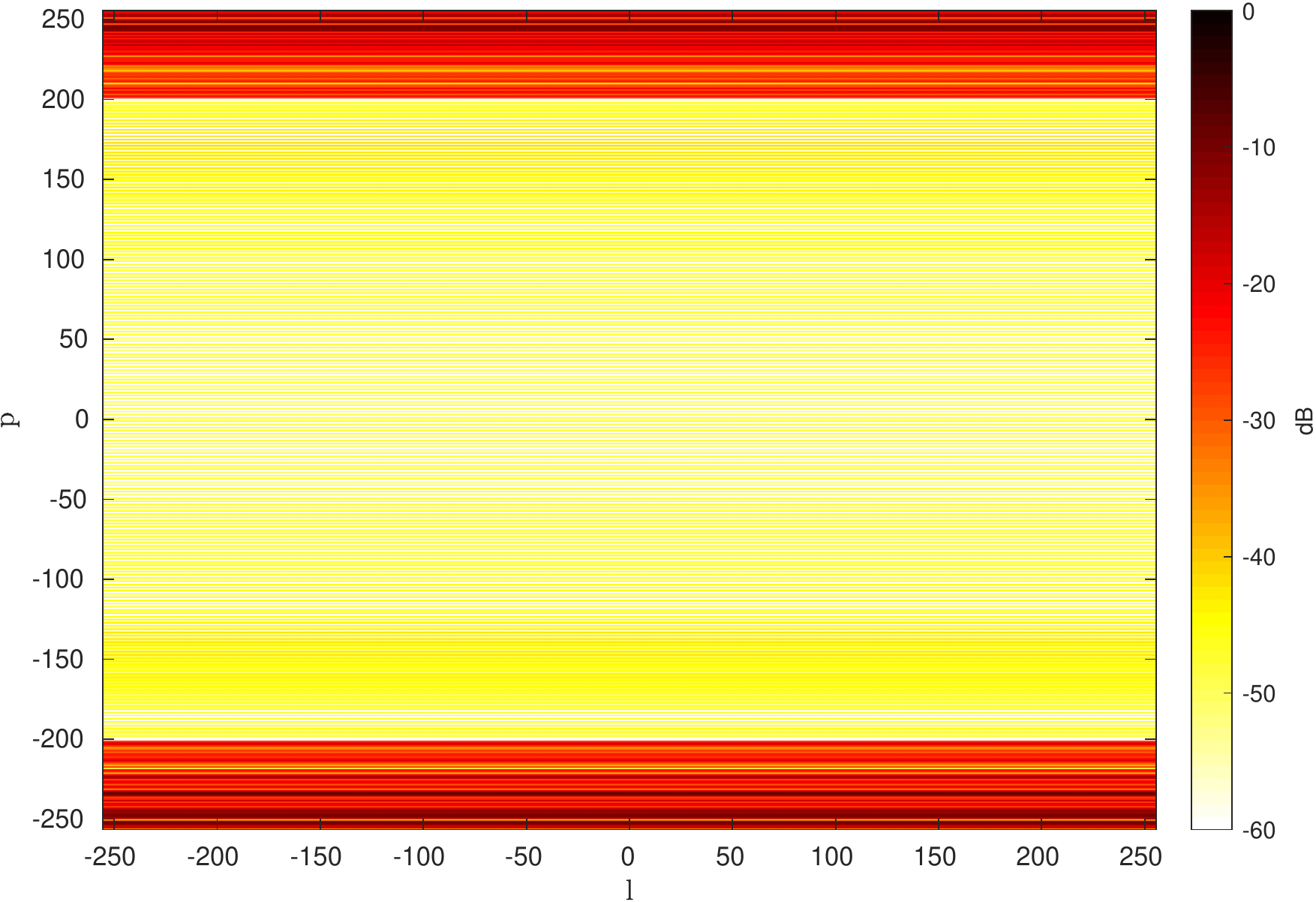}} \label{Fig:3b}} }\\
  {\subfigure[]{{\includegraphics[width = 0.35\textwidth,draft=false]{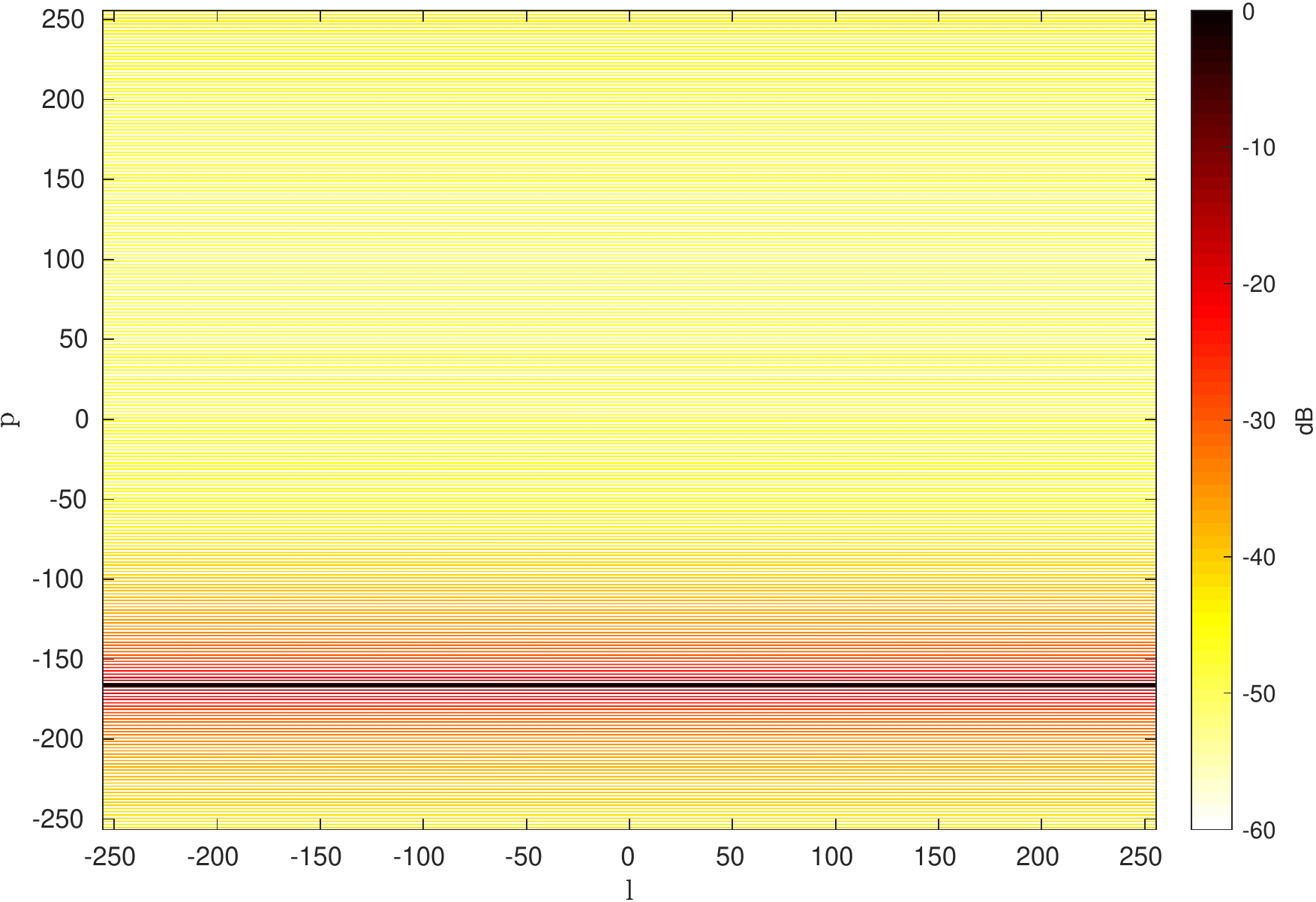}} \label{Fig:3c}} }
  {\subfigure[]{{\includegraphics[width = 0.35\textwidth,draft=false]{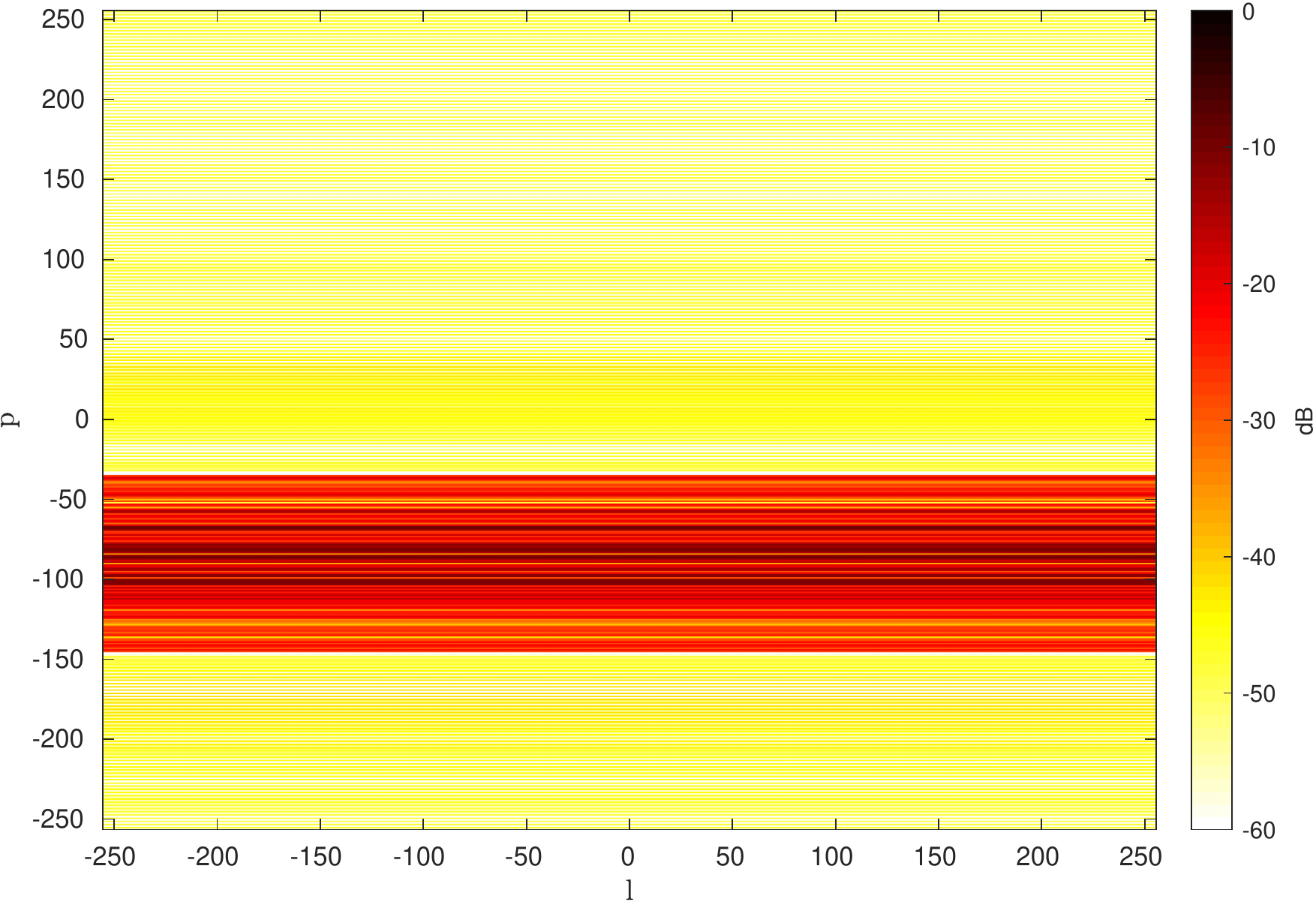}} \label{Fig:3d}} }
  \caption{Discrete periodic cross-ambiguity functions for (a) a pair of Doppler-shift codes, (b) a pair of the optimized codes $\mathbf{x}$ and $\mathbf{y}$, (c) optimized code $\mathbf{x}$ and an all-one code, and (d) an all-one code and optimized code $\mathbf{y}$, for $N=256$, $P=200$ and $N_f = 512$.}\label{Fig:3}
\end{figure}

\begin{figure}[!t]
  \centering
  \includegraphics[width = 0.6\textwidth,draft=false]{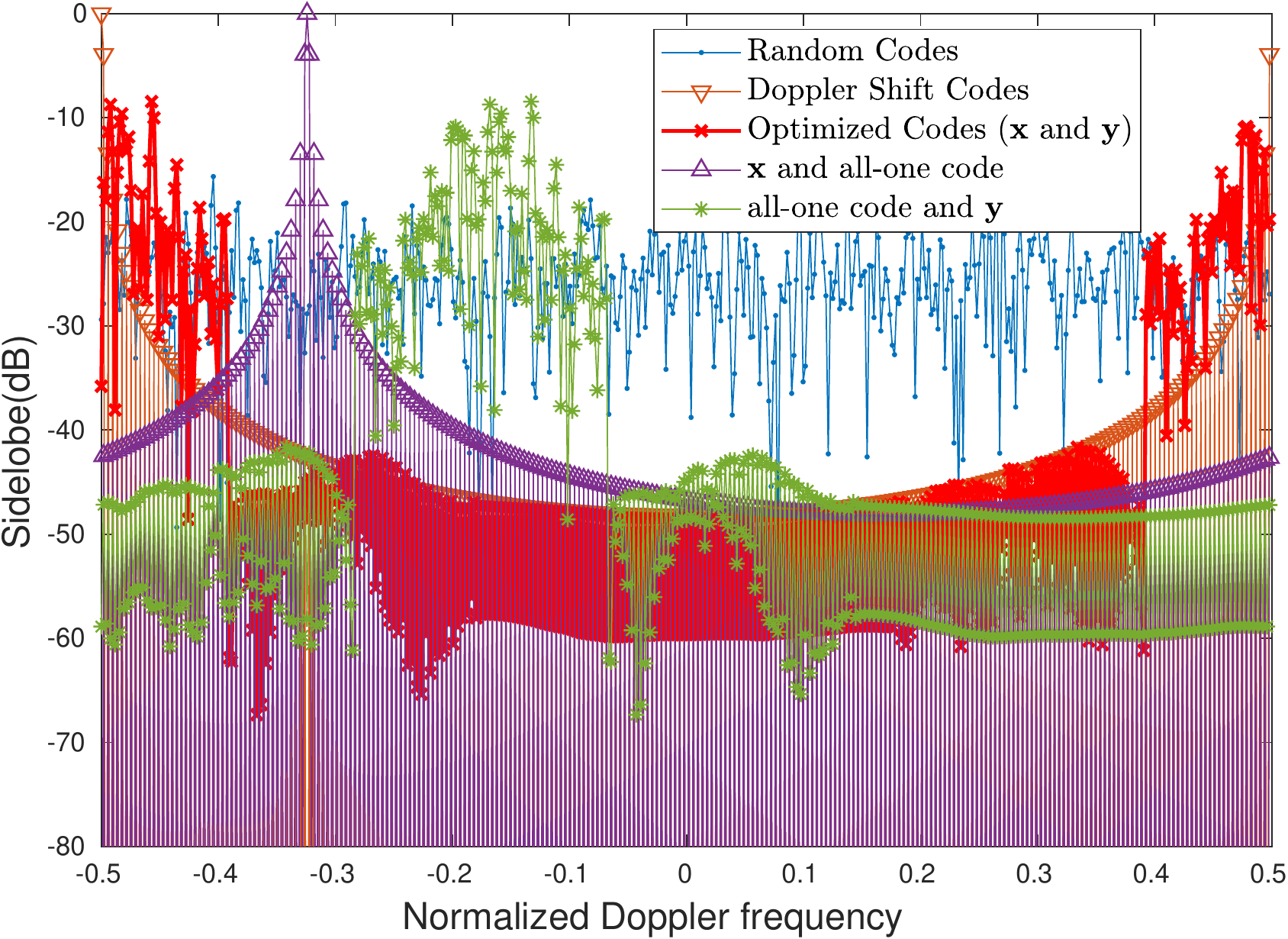}
  \caption{Comparison of the discrete periodic cross-ambiguity functions at the zero-delay cut for different scenarios for $N=256$, $P=200$ and $N_f = 512$. Although the optimized scheme achieves lower sidelobes in the desired area, the sidelobe increases in the regions outside the desired area.} \label{Fig:4}
\end{figure}

In Fig. \ref{Fig:3a} and \ref{Fig:3b}, the discrete PCAFs are shown for a pair of Doppler-shifting codes, and the optimized codes, respectively, where $P=200$ and $N_f = 512$ (which implies that the maximum Doppler frequency of interest should be lower than $3906.25$ Hz, corresponding to a maximum relative radial velocity of $87.9$ km/h).
In addition, Fig. \ref{Fig:3c} and \ref{Fig:3d} show the PCAFs when one of the vehicles uses an optimized code and the other uses a non-cooperative code such as the all-one vector. 
It is clear from the figure that a non-cooperating interfering radar present in the scenario may ultimately fail to decrease the sidelobe in the desired region.

Fig. \ref{Fig:4} compares the discrete periodic cross-ambiguity functions at the zero-delay cut for the above scenarios.
One can observe that both Doppler shifting and optimized coding schemes achieve very low sidelobes in the desired area compared to random codes.
Moreover, although the optimized scheme achieves lower sidelobes in the desired area, the sidelobe increases in the regions outside the desired area.
However, for the Doppler shift coding, the sidelobes are spread evenly throughout the entire region.
Therefore, they can be used to effectively suppress the interference.
Further note that the peak side-lobe level (PSL) corresponding to the optimized codes is approximately $3.55$ dB lower than that of the Doppler-shifting, within the desired range of Doppler frequency of interest. Interestingly, if we fix $\mby = \bone_N$ and only optimize $\mbx$, we obtain similar results, which corresponds to a scenario where no coordination between the two radar systems is required.

\begin{figure}[!t]
  \centering
  \includegraphics[width = 0.6\textwidth,draft=false]{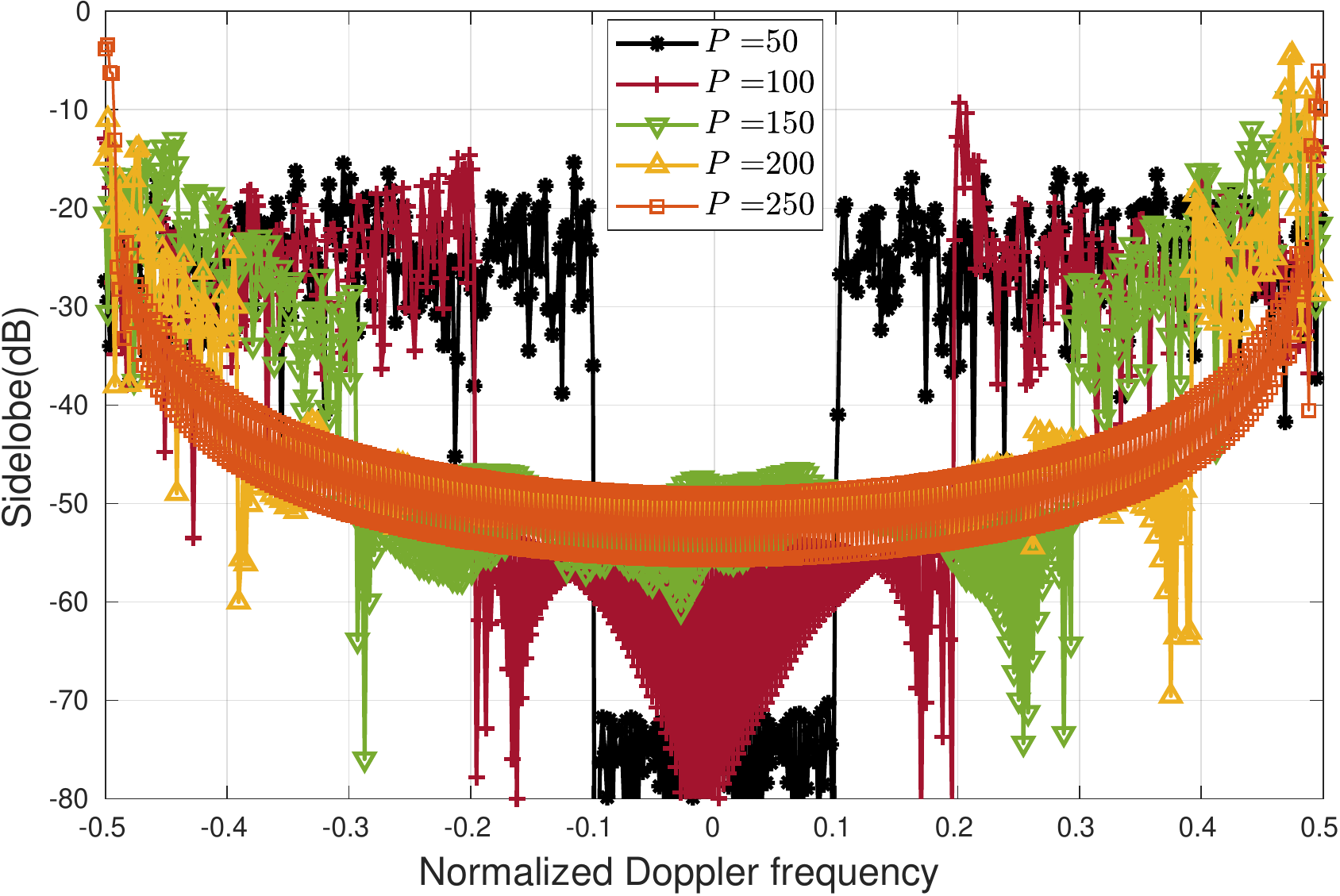}
  \caption{Comparison of the discrete PCAFs at the zero-delay cut for the optimized SISO codes with $P\in \{50, 100, 150, 200, 250\}$, $N=256$ and $N_f = 512$.} \label{Fig:66}
\end{figure}

Next, we examine the performance of the optimized SISO codes for different values of $P$. In order to preserve the fairness in this experiment, we fix the maximum outer iteration number of Algorithm \ref{Alg:1} to $1000$ for each $P$. A comparison of the sidelobe level at the zero-delay cut of the PCAFs for $P\in \{50, 100, 150, 200, 250\}$ is shown in Fig.~\ref{Fig:66}. It can be seen from the figure that the sidelobes in the Doppler frequency of interest increase with the increase in $P$. However, it is interesting to note that, once we fix $\epsilon$ in Algorithm \ref{Alg:1} instead of fixing the maximum iteration number, the sidelobe performance in the Doppler frequency of interest improves for higher $P$, however, it takes the algorithm longer to converge. 

\begin{figure}[!t]
  \centering
  {\subfigure[]{{\includegraphics[width = 0.35\textwidth,draft=false]{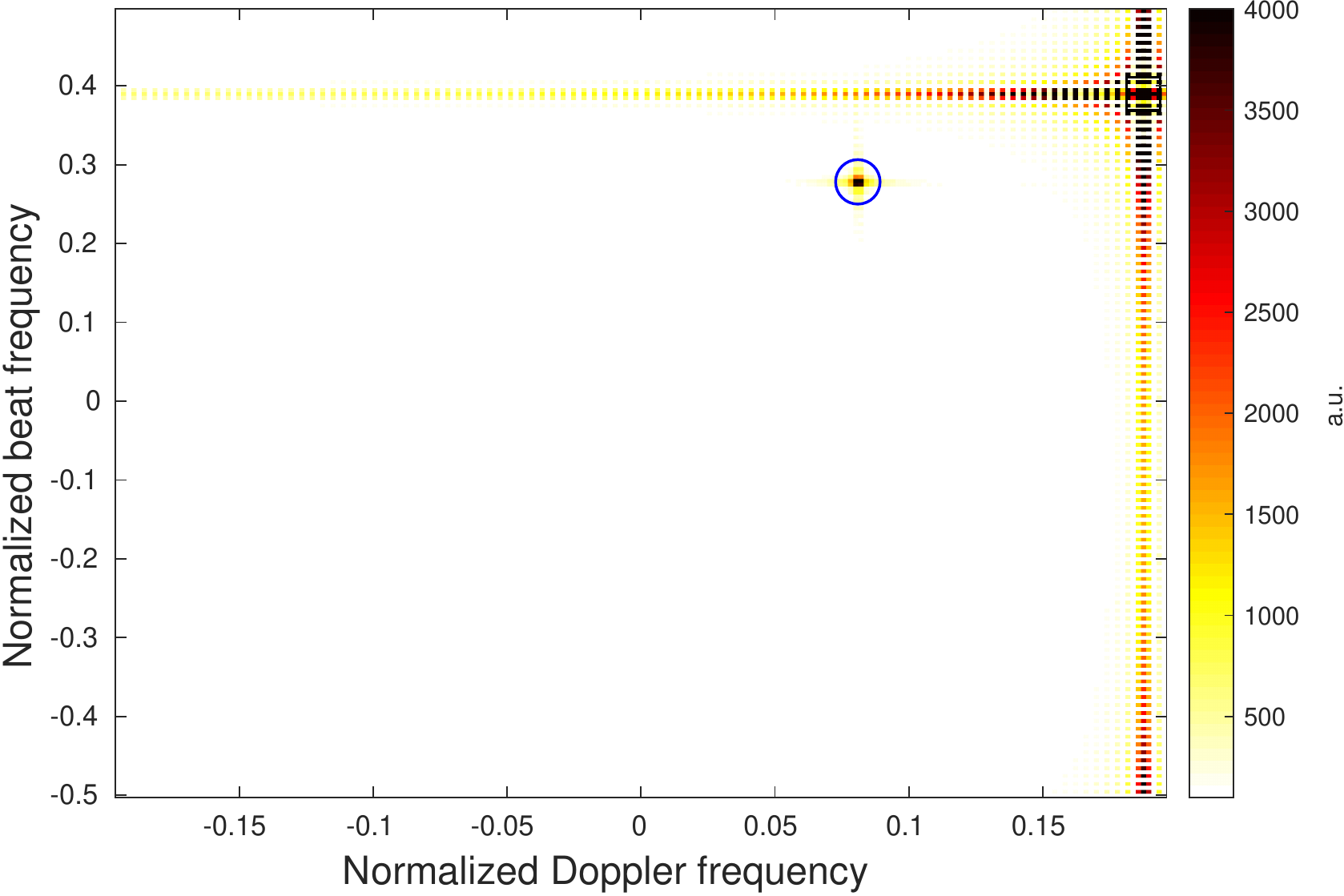}} \label{Fig:6a}} }
  {\subfigure[]{{\includegraphics[width = 0.35\textwidth,draft=false]{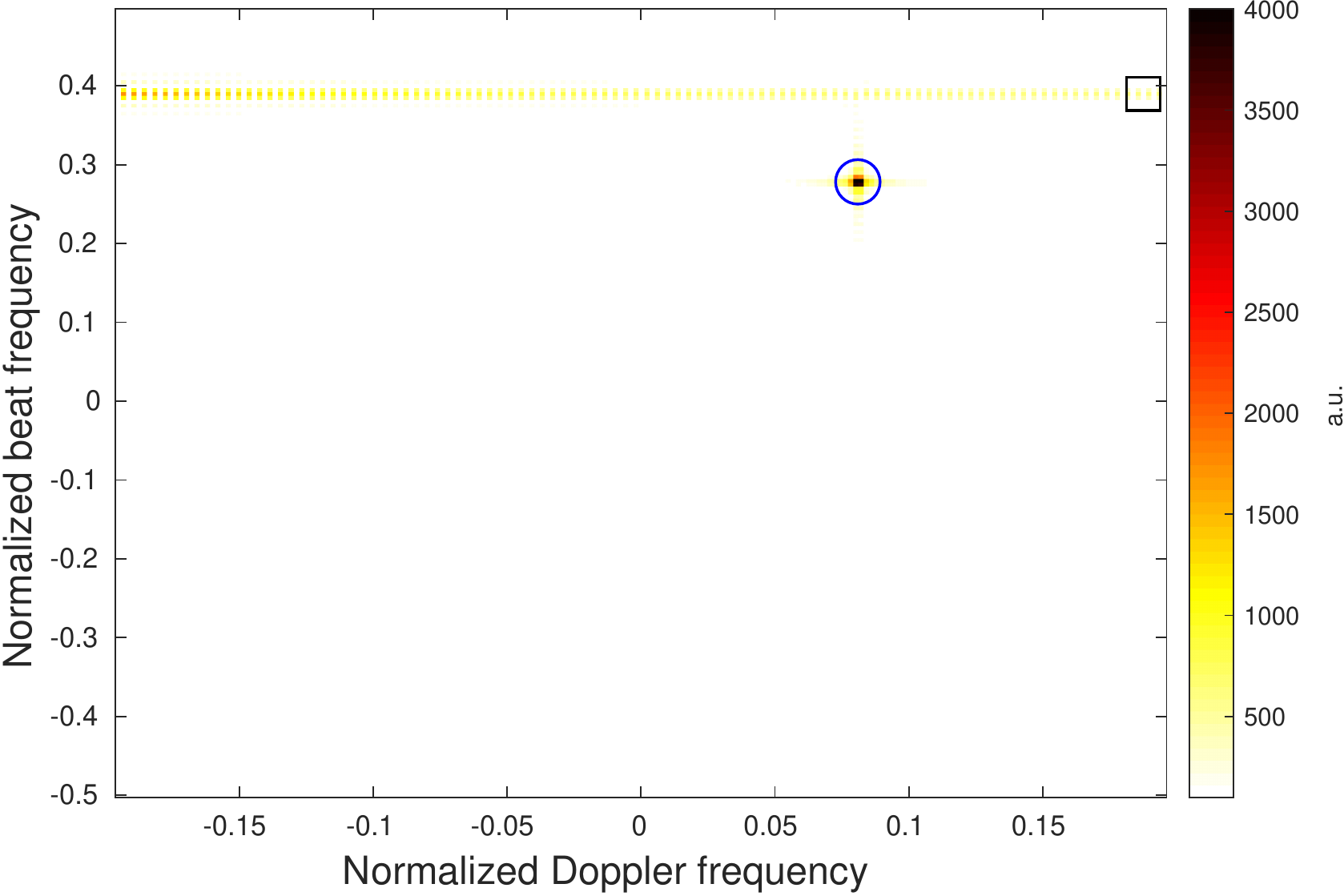}} \label{Fig:6b}} }
  {\subfigure[]{{\includegraphics[width = 0.35\textwidth,draft=false]{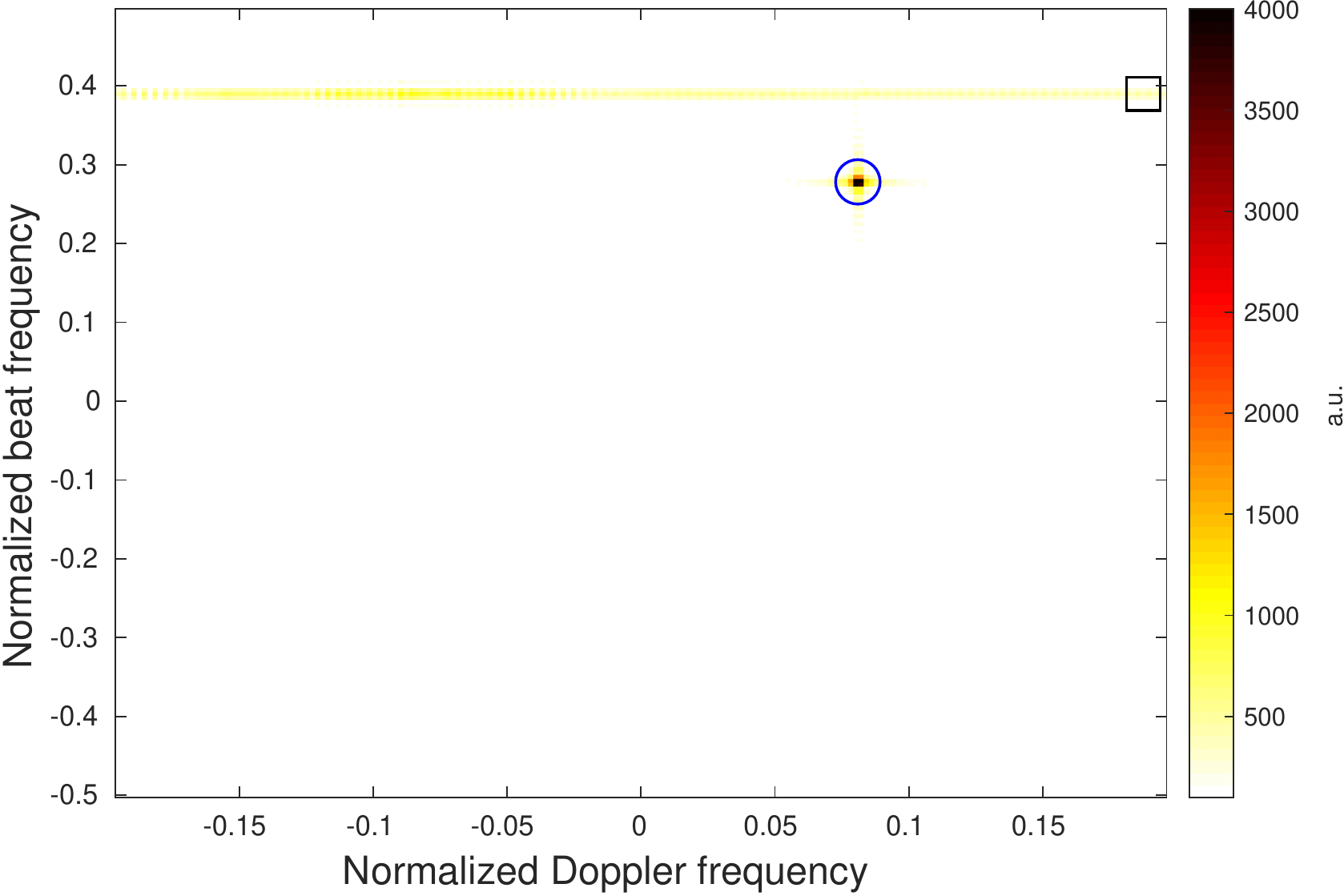}} \label{Fig:6c}} }
  {\subfigure[]{{\includegraphics[width = 0.35\textwidth,draft=false]{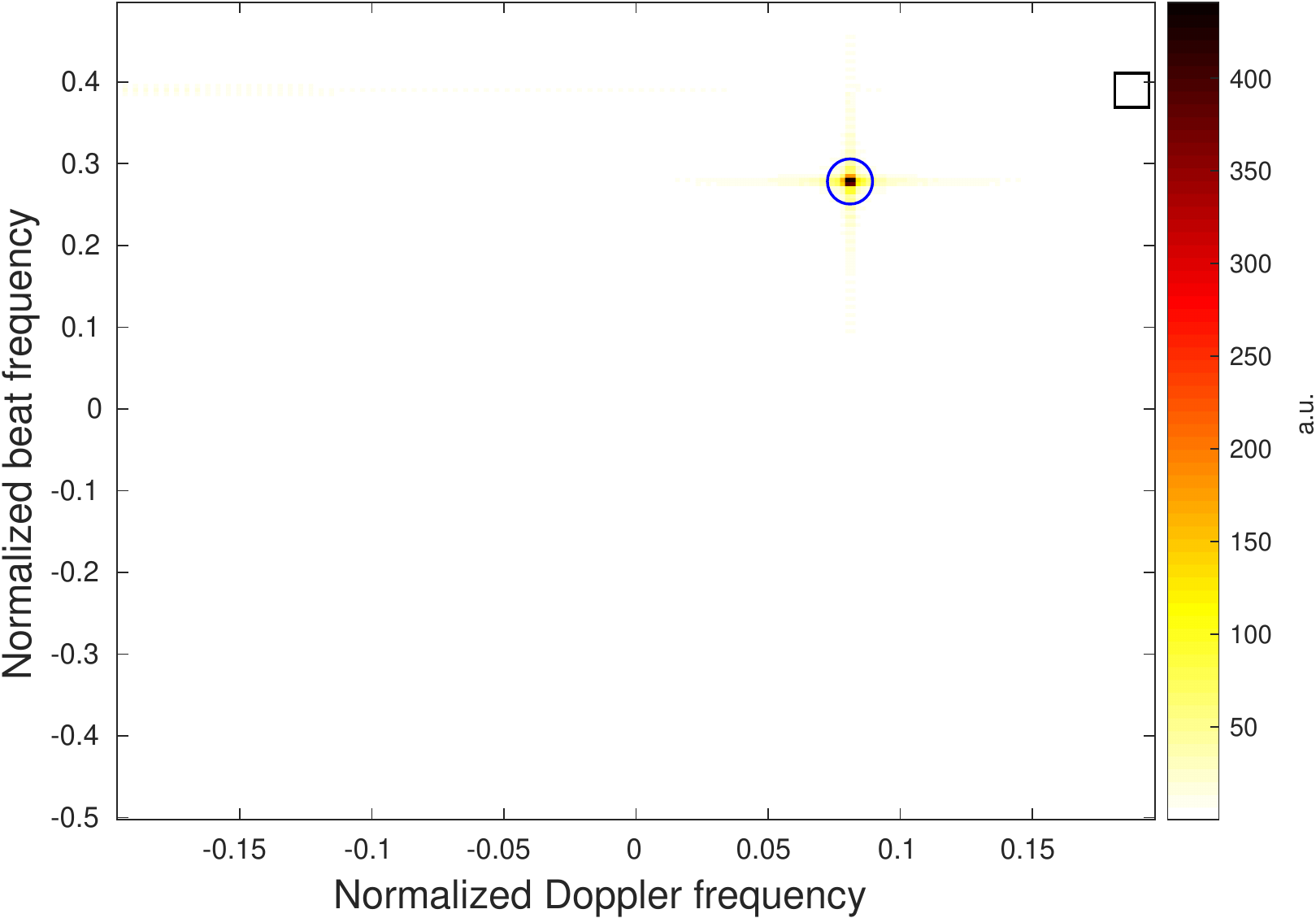}} \label{Fig:6d}} }
  {\subfigure[]{{\includegraphics[width = 0.35\textwidth,draft=false]{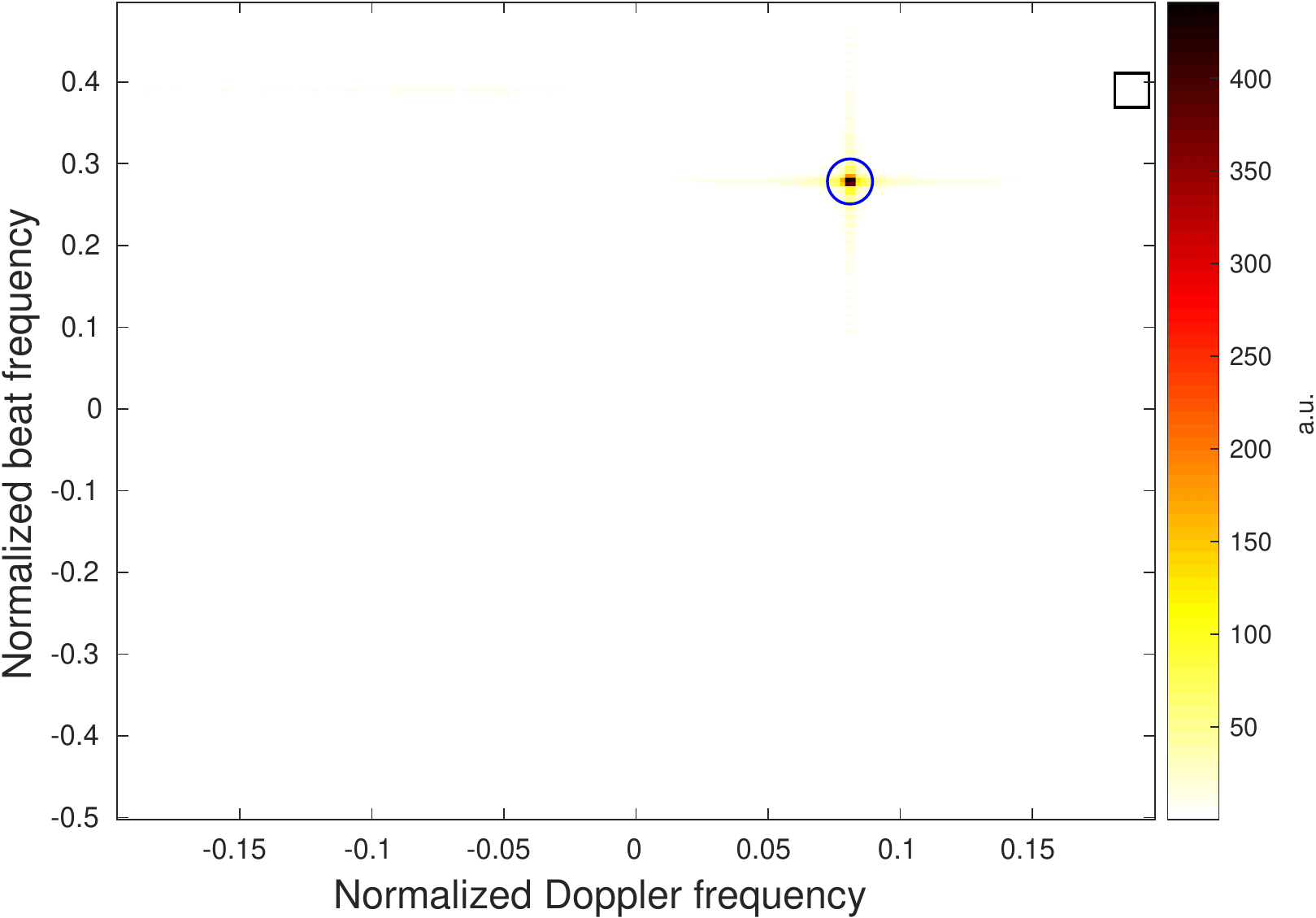}} \label{Fig:6e}} }
  \caption{The range-Doppler image for (a) randomly generated code without slow-time coding, (b) Doppler-shift coding, and (c) the optimized coding scheme for the SNRs of target and interferer as 30 dB and 60 dB,  respectively. Furthermore, the same for (d) Doppler-shift coding, and (e) the optimized coding scheme for the SNRs of target and interferer as 10 dB and 20 dB,  respectively: \blue{$\bigcirc$} represents the target and $\Box$ represents the interference.}\label{Fig:6}
\end{figure}

We, next, apply these coding schemes to mitigate the mutual interference for two identical automotive radar systems operating in a typical scenario: The range of target and interference are at $50$ m and $70$ m, respectively. The speeds associated with them are $10.12$ m/s and $23.45$ m/s. The signal-to-noise ratios (SNR) are $30$ dB and $60$ dB, respectively. The sampling frequency is $f_s = 4$ MHz. $M = 100$ samples are collected for each period.
Fig.~\ref{Fig:6} (a-c) show the range-Doppler image in this scenario without slow-time coding, using Doppler-shifting code, and the optimized coding scheme, respectively. Furthermore, Fig.~\ref{Fig:6} (d-e) show the same for Doppler-shifting code and the optimized coding scheme when the SNRs are $10$ dB and $20$ dB for the target and interferer, respectively. We can observe that the power of the interference is much stronger than that of the target such that a false alarm occurs. When our slow-time coding schemes are applied, the interference power level is significantly reduced and the target can be easily detected without suffering from false alarm problems.

\subsection{The MIMO Scenario}
\begin{figure}[!t]
	\centering
	\hspace{-.5cm}
	\includegraphics[width=0.6\textwidth,draft=false]{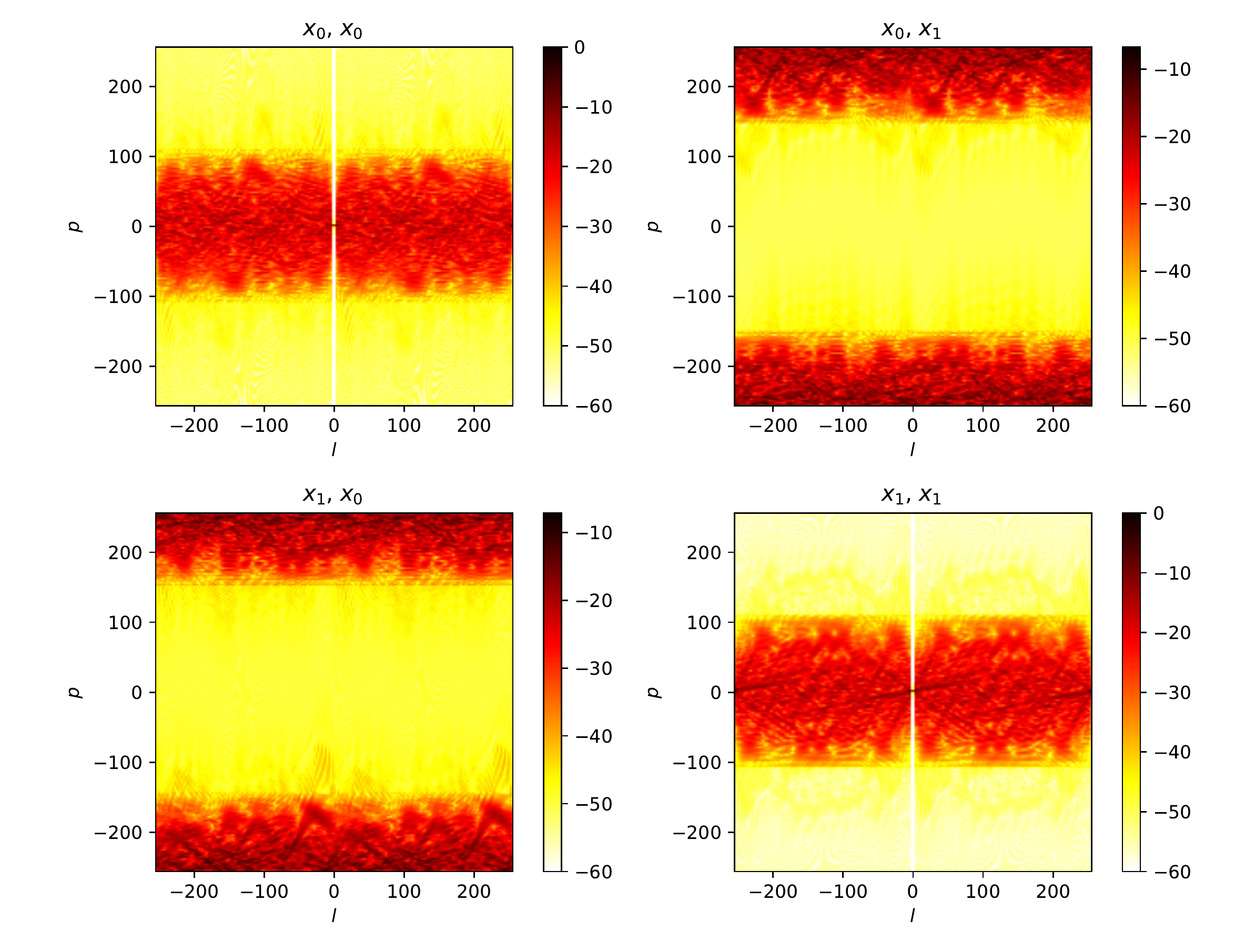}
	\caption{Discrete periodic (cross) ambiguity functions between different MIMO sequences from the set $\mathcal{X}=\{\mathbf{x}_m\}_{m=1}^2$ for $N=256$, $P=200$ and $N_f = 512$. Note that for identical sequences, the ambiguity function always assumes larger values near zero.}
	\label{fig:mimo_X}
\end{figure}
\begin{figure}[!t]
	\centering
	\hspace{-.3cm}
	\includegraphics[width=0.8\textwidth,draft=false]{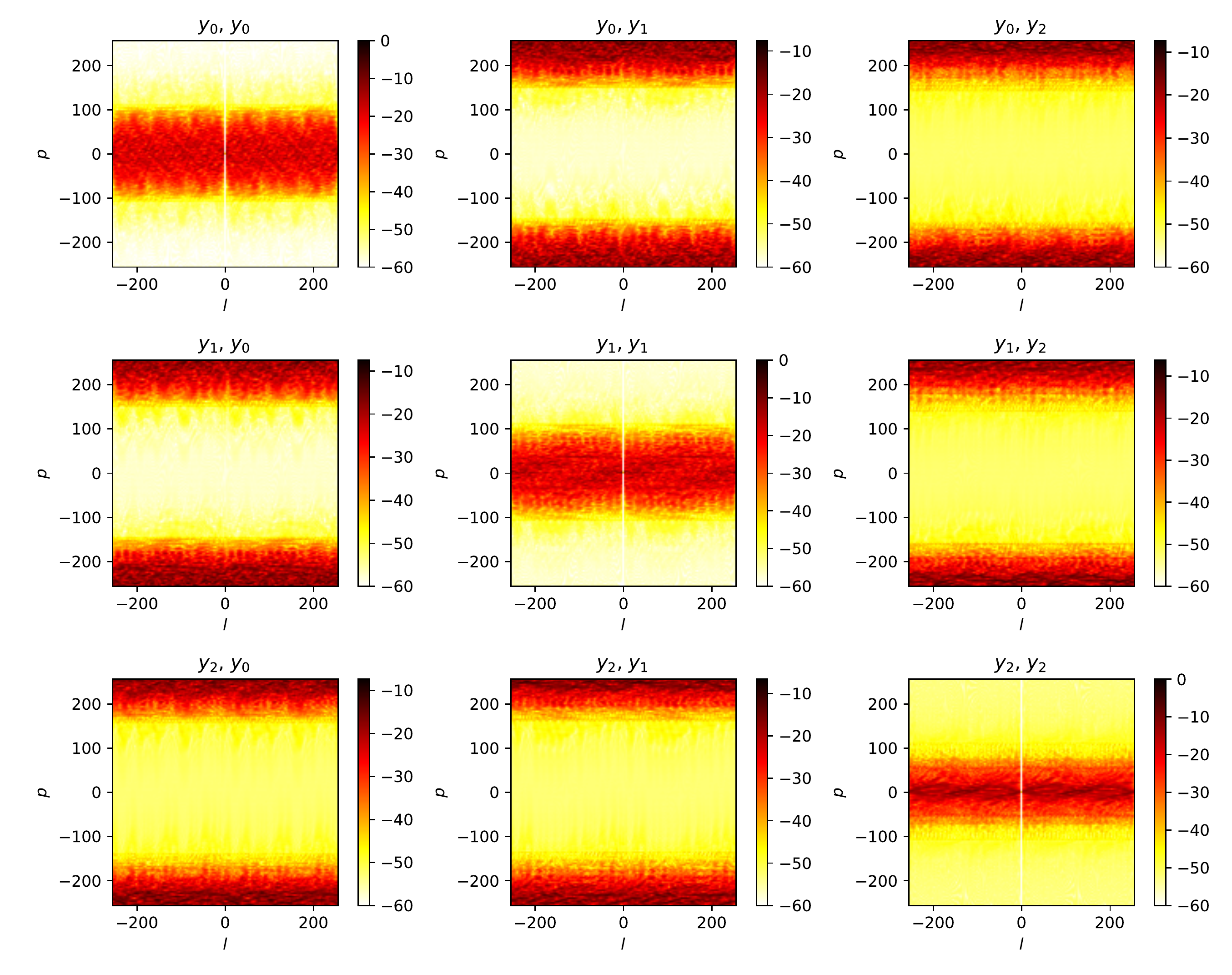}
	\caption{Discrete periodic (cross) ambiguity functions among different MIMO sequences from the set $\mathcal{Y}=\{\mathbf{y}_k\}_{k=1}^3$ for $N=256$, $P=200$ and $N_f = 512$.}
	\label{fig:mimo_Y}
\end{figure}
\begin{figure}[!t]
	\centering
	\hspace{-.3cm}
	\includegraphics[width=0.8\textwidth,draft=false]{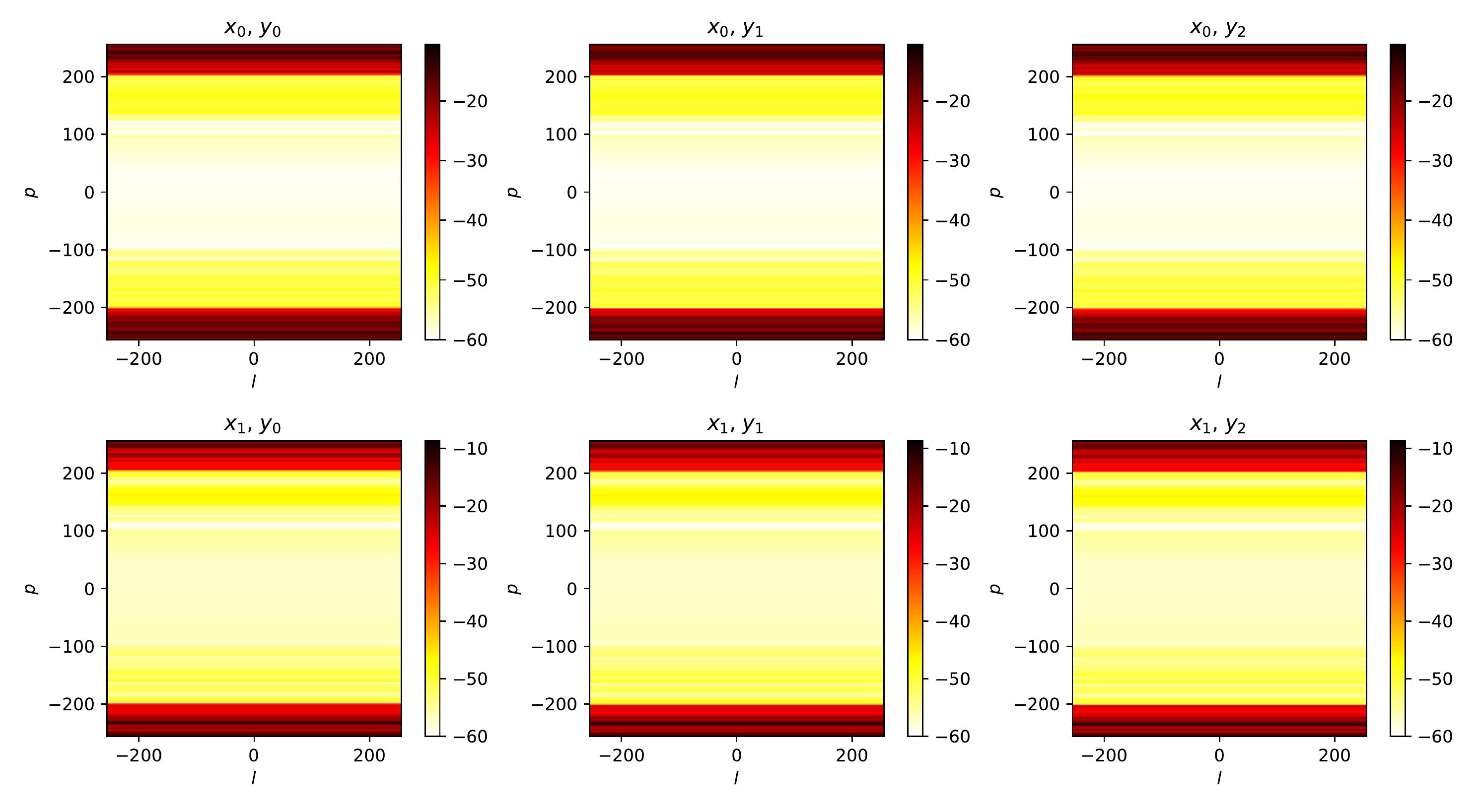}
	\caption{Discrete periodic cross-ambiguity functions between different MIMO sequences from the sets $\mathcal{X}=\{\mathbf{x}_m\}_{m=1}^2$ and $\mathcal{Y}=\{\mathbf{y}_k\}_{k=1}^3$ for $N=256$, $P=200$ and $N_f = 512$.}
	\label{fig:mimo_XY}
\end{figure}
In order to examine the effectiveness of Algorithm \ref{Alg:2} in the MIMO case, we use a similar set of FMCW parameters as in the SISO case.
For the ease of display and sake of simplicity, we consider two similar MIMO FMCW radar systems that use set of codes $\mathcal{X}=\{\mathbf{x}_m\}_{m=1}^2$ and $\mathcal{Y}=\{\mathbf{y}_k\}_{k=1}^3$  each of length $N=256$, operating under the same carrier frequency of $f_c = 24$ GHz, the bandwidth of the chirp signal of $B = 150$ MHz, and the sweep time of $T_c = 50$ $\mu$s.
As mentioned earlier, we first initialize the algorithm with normally distributed random numbers for all optimization variables: $\{\mathbf{x}_m\}$, $\{\mathbf{y}_k\}$ $\{\mathbf{u}^{c}_{l,p,m}\}$, $\{\mathbf{v}^{c}_{l,p,k}\}$, for  $c \in \{r,i\}$. 
After running the algorithm once, we use the output codes $\{\mathbf{x}_m\}$, $\{\mathbf{y}_k\}$ as the initial codes for the next run but we still use random initialization for other variables, $\{\mathbf{u}^{c}_{l,p,m}\}$, $\{\mathbf{v}^{c}_{l,p,k}\}$, and repeat the process multiple times.

Figs. \ref{fig:mimo_X}-\ref{fig:mimo_XY} show the periodic (cross) ambiguity functions for the radar systems: $\{\mathcal{X}\}, \{\mathcal{Y}\}$, and $\{\mathcal{X}, \mathcal{Y}\}$ where $P=200$ and $N_f=512$. It is evident from each of these figures that for each set of sequences, the unambiguous regions are well separated (within the range of -40dB to -60dB), and hence, these sequences can be reliably used in MIMO FMCW radar systems that require mutual interference mitigation.

\begin{figure}[!t]
  \centering
  {\subfigure[]{{\includegraphics[width = 0.35\textwidth,draft=false]{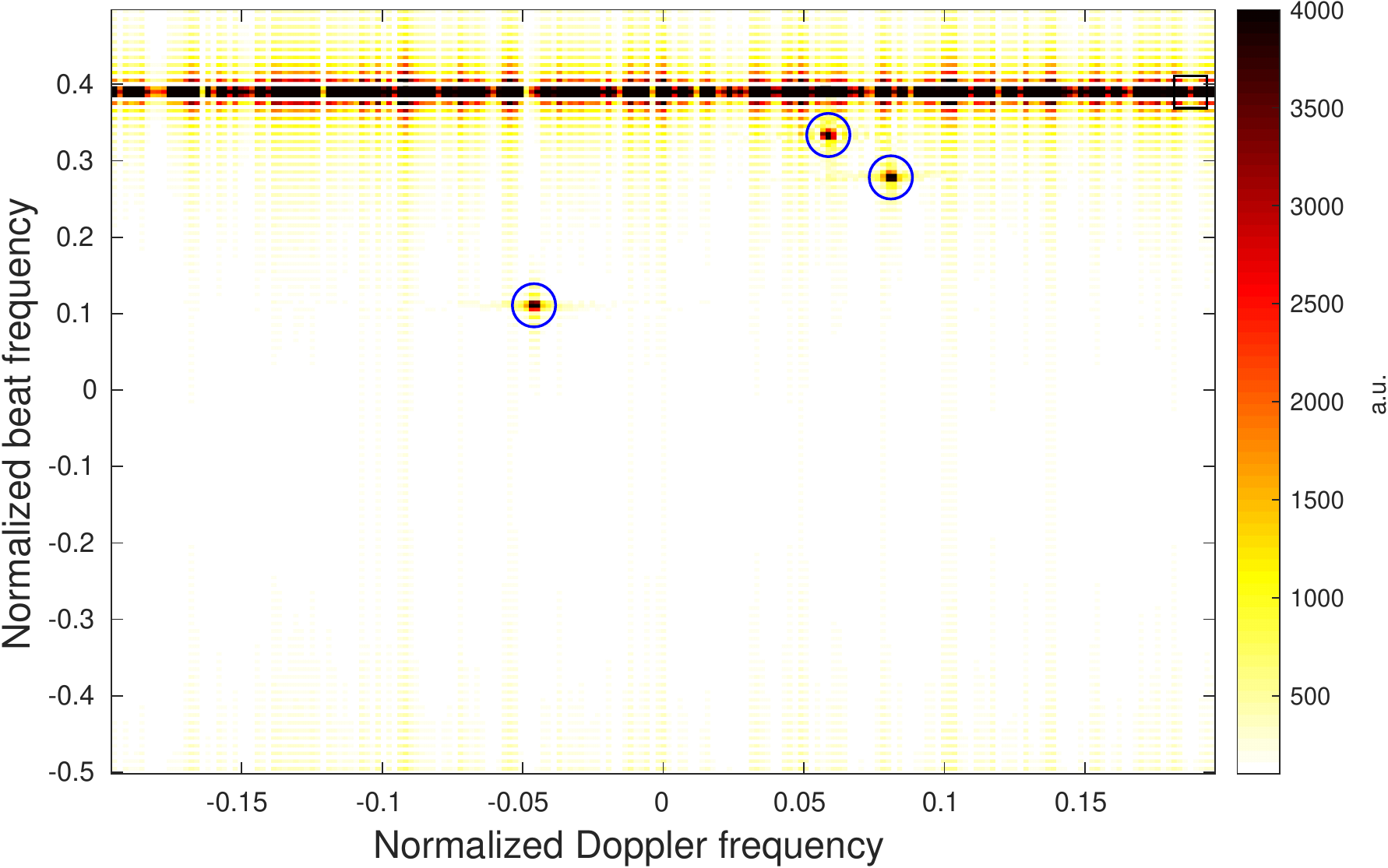}} \label{Fig:11a}} }
  {\subfigure[]{{\includegraphics[width = 0.35\textwidth,draft=false]{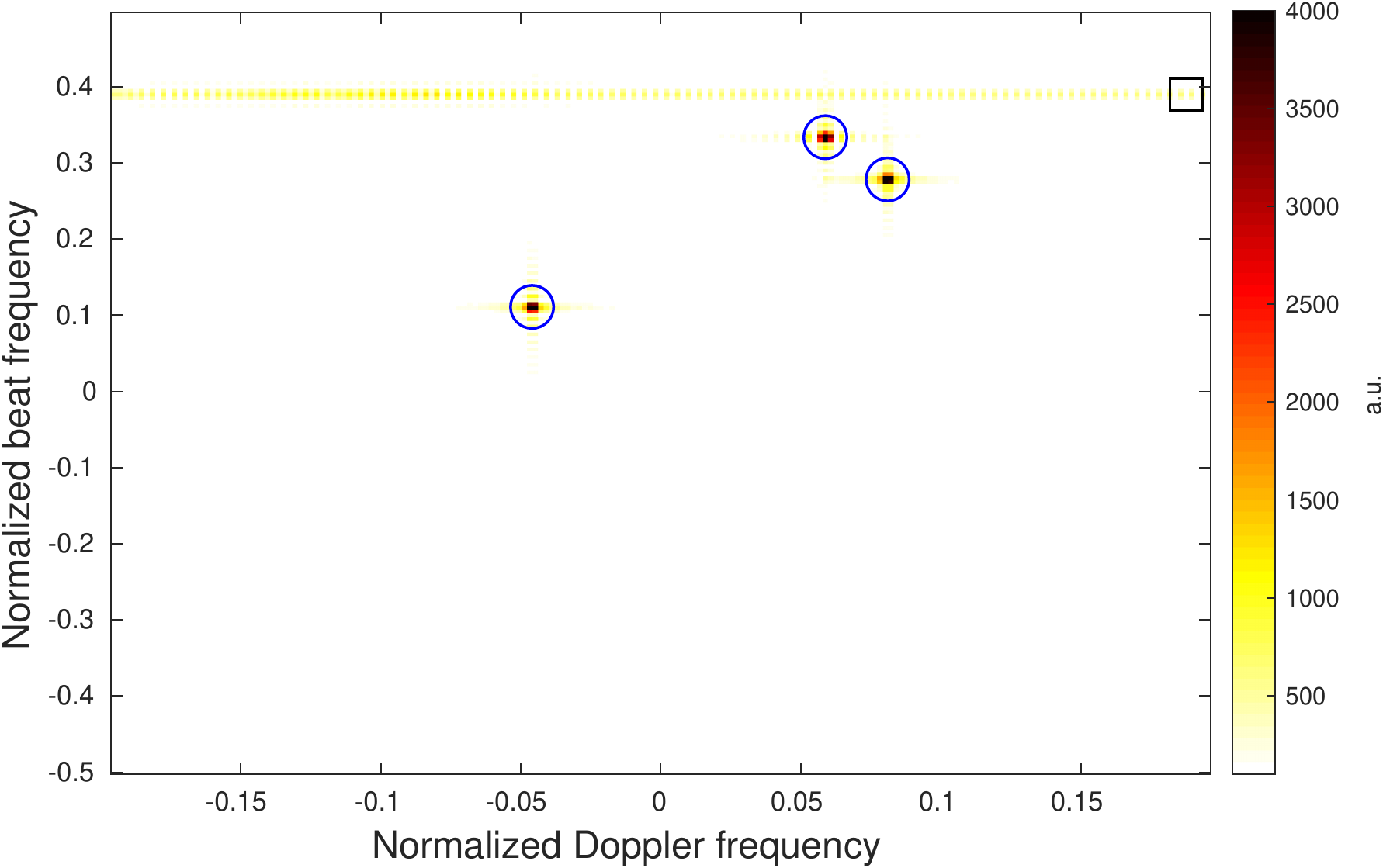}} \label{Fig:11b}} }
  {\subfigure[]{{\includegraphics[width = 0.35\textwidth,draft=false]{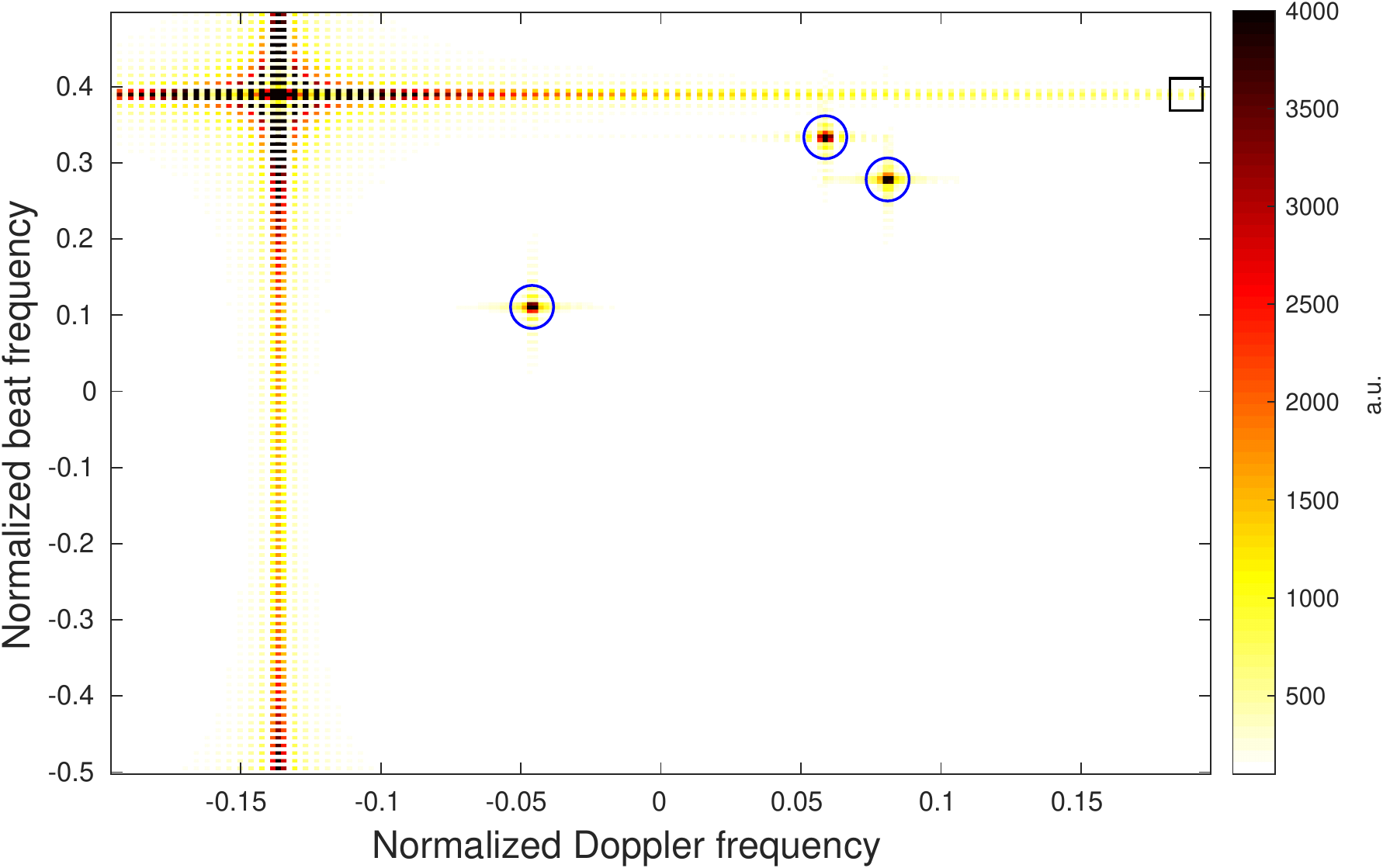}} \label{Fig:11c}} }
  \caption{The range-Doppler image for (a) randomly generated MIMO codes without slow-time coding, (b) the optimized MIMO coding scheme, and (c) optimized MIMO scheme, however the interferer uses a non-cooperating all-one vector code: \blue{$\bigcirc$} represents the targets and $\Box$ represents the interference.}\label{Fig:11}
\end{figure}

In the next example for the MIMO case, we apply the optimized coding scheme to mitigate the mutual interference in the presence of multiple targets. For this scenario, we use three targets and one interfering radar system. The ranges of the targets and interference are at $50, 20, 60$ m, and $70$ m, respectively. The speeds associated with them are $10.12, -5.75, 7.34$ m/s and $23.45$ m/s. The SNRs are $30$ dB and $60$ dB for the targets and interference, respectively. The sampling frequency is similarly $f_s = 4$ MHz as assumed in the SISO case.
Fig.~\ref{Fig:11} shows the range-Doppler (RD) image for different scenarios.
The RD images are shown for randomly generated MIMO codes without slow-time coding, and the optimized MIMO coding scheme in Fig.~\ref{Fig:11a} and ~\ref{Fig:11b}, respectively.
It is clear from the figure that the interference power level for the optimized codes is significantly reduced and all three of the targets are easily detected without suffering from false alarm issues.
Furthermore, Fig.~\ref{Fig:11c} shows the RD image when the interferer uses a non-cooperating all-one vector code as described in the SISO case.
It is interesting to see that the presence of a strong interferer that does not respect the code usage protocol may significantly hinder the target detection capabilities.

\section{Conclusion} \label{sec:con}
In this paper, we discussed the problem of mutual interference mitigation in identical or similar radar systems employed in automotive applications. 
We proposed two slow-time coding schemes for the SISO case. Specifically, the first coding scheme is designed through Doppler shifting and the second one is devised based on an efficient cyclic optimization approach. We further extended the problem formulation and proposed another efficient algorithm to generate the radar codes in the more general MIMO scenario. We showed that these coding schemes can be used to reduce the interference power level significantly. 
Note that these coding schemes are highly effective when the radar parameters are nearly identical, however, the requirement of having identical parameters is not necessary.

\appendices
\section{Efficient Computation of $\mbB_x$ and $\mbB_y$}\label{app:apdA} 

Note that $\mbB_y$ in \eqref{Eq:By} can be rewritten as
\begin{equation}\label{Eq:By_Interest}
  \mbB_y = \sum_{p=-P}^{P} \Diag{\mbf_p} \left(\sum_{l=-N+1}^{N-1} \mbC_l \mby \mby^H \mbC_l^H\right) \Diag{\mbf_p}^H.
\end{equation}
Define $\mbR = \sum_{l=-N+1}^{N-1} \mbC_l \mby \mby^H \mbC_l^H$. It is easy to verify  $\mbR_y = 2\mbR_0 - \mby\mby^H$, considering that for $l<0$, $\mbC_l \mby = \mbC_{N-l} \mby$, and $\mbR_0 = \sum_{l=0}^{N-1} \mbC_l \mby \mby^H \mbC_l^H$. Moreover, we can write $\mbR_0$ as
\begin{equation}
  \mbR_0 = \mbY\mbY^H,
\end{equation}
where
\begin{equation}
  \mbY = \begin{bmatrix}
            y_1 & y_2 & \cdots & y_{N-1} & y_N \\
            y_2 & y_3 & \cdots & y_N & y_1 \\
            \vdots & \vdots & \cdots & \vdots & \vdots \\
            y_N & y_1  & \cdots & y_{N-2} & y_{N-1} \\
          \end{bmatrix}.
\end{equation}
As a result, $\mbR_0$ can be written as
\begin{equation*}
  \mbR_0 = \begin{bmatrix}
            c_{y,0} & c_{y,1} & \cdots & c_{y,N-2} & c_{y,N-1} \\
            c_{y,1}^{*} & c_{y,0} & \cdots & c_{y,N-3} & c_{y,N-2} \\
            \vdots & \vdots & \cdots & \vdots & \vdots \\
            c_{y,N-1}^* & c_{y,N-2}^*  & \cdots & c_{y,1}^* & c_{y,0} \\
          \end{bmatrix},
\end{equation*}
where $c_{y,l} = \sum_{n=1}^N y_n y^*_{(n+l)\textrm{mod} N}, l=0,1,\cdots,N-1$. It should be noted that $\mbR_0$ is a circulant matrix and its elements are determined by the values of the sequence $\{c_{y,l}\}_{l=0}^{N-1}$. Moreover, this sequence can be seen as the circular convolution between  $\{y_n\}_{n=1}^N$ and $\{y_n\}_{n=1}^N$, and hence can be efficiently computed  via FFT operations. Therefore, the computation of $\mbR_y$ requires $\Oh{N^2}$ flops (mainly due to the computation of $\mby\mby^H$).

It follows that $\mbB_y$ can be calculated efficiently using the following result:
\begin{align}\label{Eq:By_Interest2}
  \mbB_y &= \sum_{p=-P}^{P} \Diag{\mbf_p} \bR_y \Diag{\mbf_p}^H \nonumber\\
        &= \sum_{p=-P}^{P} \mbR_y \odot (\mbf_p \mbf_p^H) \nonumber\\
        & = \mbR_y \odot (\mbF_P \mbF_P^H),
\end{align}
where $\mbF_P = [\mbf_{-P}, \cdots,\mbf_P] \in \mathbb{C}^{N\times (2P+1)}$.

Finally,  we consider reducing the computational complexity of computing $\mbB_x$. To this end, we note that
\begin{align}\label{Eq:BxFast}
  \mbB_x &= \sum_{l=-N+1}^{N-1}\mbC_l^H \left(\sum_{p=-P}^{P}  \Diag{\mbf_p}^H\mbx\mbx^H\Diag{\mbf_p}\right) \mbC_l \nonumber\\
        &= \sum_{l=-N+1}^{N-1}\mbC_l^H\left((\mbx\mbx^H) \odot (\mbF_P^* \mbF^T_P)  \right)\mbC_l.
\end{align}
Let $\mbR_x = (\mbx\mbx^H) \odot (\mbF_P^* \mbF^T_P)$. We can observe that $\mbC_l^H \mbR_x \mbC_l$ can be computed very efficiently, since it only involves the permutation of the rows and columns of $\mbR_x$. Therefore, the computation of $\mbB_y$ only requires $\Oh{N^2}$ flops.

\bibliographystyle{IEEEbib}
\bibliography{refs}
\balance
\end{document}